\documentclass[twocolumn,epjc3]{svjour3}  

\RequirePackage{graphicx}
\RequirePackage{lineno}
\usepackage{multirow}

\journalname{Eur. Phys. J. C}

\begin{document}

\title{Pulse shape analysis in {\sc Gerda} Phase~II}

\titlerunning{Pulse shape analysis in {\sc Gerda} Phase~II}

\author{
The \mbox{\protect{\sc{Gerda}}} collaboration\thanksref{ALNGS,corrauthor}
  \\[4mm]
M.~Agostini\thanksref{UCL,TUM} \and
G.~Araujo\thanksref{UZH} \and
A.M.~Bakalyarov\thanksref{KU} \and
M.~Balata\thanksref{ALNGS} \and
I.~Barabanov\thanksref{INRM} \and
L.~Baudis\thanksref{UZH} \and
C.~Bauer\thanksref{HD} \and
E.~Bellotti\thanksref{MIBF,MIBINFN,dec} \and
S.~Belogurov\thanksref{ITEP,INRM,alsoMEPHI} \and
A.~Bettini\thanksref{PDUNI,PDINFN}
% ...
}
\authorrunning{the \textsc{Gerda} collaboration}
\thankstext{corrauthor}{
  \emph{correspondence:}  gerda-eb@mpi-hd.mpg.de}
\thankstext{dec}{deceased}
\thankstext{alsoMEPHI}{\emph{also at:} NRNU MEPhI, Moscow, Russia}
\thankstext{nowKIT}{\emph{present address:} Institut f{\"u}r  Experimentelle
  Teilchenphysik,  Karlsruher Institut f{\"u}r Technologie, Karlsruhe , Germany} 
\thankstext{alsoLev}{\emph{also at:} Moscow Inst. of Physics and Technology,
  Russia}
\thankstext{nowBosch}{\emph{present address:} Robert Bosch GmbH, Stuttgart,
  Germany}
\thankstext{alsoDubna}{\emph{also at:} Dubna State University, Dubna, Russia} 
\thankstext{nowBerkeley}{\emph{present address:} Nuclear Science Division, Berkeley, USA}
\thankstext{nowTUM}{\emph{present address:} Physik Department, Technische  Universit{\"a}t M{\"u}nchen, Germany} 
\institute{
INFN Laboratori Nazionali del Gran Sasso and Gran Sasso Science Institute, Assergi, Italy\label{ALNGS}
% ...
\and
Max-Planck-Institut f{\"u}r Kernphysik, Heidelberg, Germany\label{HD} \and
Department of Physics and Astronomy, University College London, London, UK\label{UCL} \and
Dipartimento di Fisica, Universit{\`a} Milano Bicocca, Milan, Italy\label{MIBF} \and
INFN Milano Bicocca, Milan, Italy\label{MIBINFN}
% ...
\and
Institute for Nuclear Research of the Russian Academy of Sciences, Moscow, Russia\label{INRM} \and
Institute for Theoretical and Experimental Physics, NRC ``Kurchatov Institute'', Moscow, Russia\label{ITEP} \and
National Research Centre ``Kurchatov Institute'', Moscow, Russia\label{KU}
% ...
\and
Physik Department, Technische  Universit{\"a}t M{\"u}nchen, Germany\label{TUM} \and
Dipartimento di Fisica e Astronomia, Universit{\`a} degli Studi di 
Padova, Padua, Italy\label{PDUNI} \and
INFN  Padova, Padua, Italy\label{PDINFN}
% ...
\and
Physik-Institut, Universit{\"a}t Z{\"u}rich, Z{u}rich, Switzerland\label{UZH}
}

\date{Received: date / Accepted: date}

\maketitle


\begin{abstract}

The GERmanium Detector Array ({\sc Gerda}) collaboration searched for neutrinoless double-$\beta$ decay in $^{76}$Ge using isotopically enriched high purity germanium detectors at the Laboratori Nazionali del Gran Sasso of INFN.
After Phase~I (2011-2013), the experiment benefited from several upgrades, including an additional active veto based on LAr instrumentation and a significant increase of mass by point-contact germanium detectors that improved the half-life sensitivity of Phase~II (2015-2019) by an order of magnitude.
At the core of the background mitigation strategy, the analysis of the time profile of individual pulses provides a powerful topological discrimination of signal-like and background-like events.
Data from regular $^{228}$Th calibrations and physics data were both considered in the evaluation of the pulse shape discrimination performance.
In this work, we describe the various methods applied to the data collected in {\sc Gerda} Phase II corresponding to an exposure of 103.7~kg$\cdot$yr.
These methods suppress the background by a factor of about 5 in the region of interest around $Q_{\beta\beta}= 2039$~keV, while preserving $(81\pm 3)$\% of the signal.
In addition, an exhaustive list of parameters is provided which were used in the final data analysis.



\end{abstract}


\section{Introduction}
\label{sec:intro}

Neutrinoless double-$\beta$ ($0\nu\beta\beta$) decay is a hypothetical process in which two neutrons in a nucleus are transformed simultaneously into two protons with the emission of only two electrons.
Such a process violates lepton number conservation and requires the neutrino to be its own antiparticle (Majorana particle). 
In combination with cosmological observations and direct neutrino mass measurement, a non-zero $0\nu\beta\beta$ decay rate would highly constrain the standard light, left-handed neutrino exchange mechanism via the effective Majorana neutrino mass or shed light on alternative processes \cite{Cirigliano:2018yza}.

The highest half-life sensitivity to $0\nu\beta\beta$ decay requires the experiments to achieve large target mass, high detection efficiency, good energy resolution and most complete elimination of background at the $Q$-value of the decay ($Q_{\beta\beta}$).
The goal of the GERmanium Detector Array ({\sc Gerda}) experiment was to realize a background-free\footnote{Number of expected background events at full exposure in the region of interest below 1.} experiment for the first time.
{\sc Gerda} was located at the underground Laboratori Nazionali del Gran Sasso (LNGS) of INFN, Italy.
{\sc Gerda} used up to 43~kg of high purity germanium (HPGe) detectors enriched in the candidate isotope $^{76}$Ge up to 88\%. 
They ensure high detection efficiency, low intrinsic background and excellent energy resolution. 
The bare HPGe detectors were operated in liquid argon (LAr), which served as cooling medium and as active shield against environmental backgrounds at $Q_{\beta\beta} = 2039$~keV. 
The details of the experimental setup and its upgrade from Phase~I to Phase~II can be found elsewhere~\cite{Ackermann:2012xja,Agostini:2017hit}. 

The Phase~II data taking took place between December 2015 and November 2019 with an upgrade of the detector array and the surrounding LAr instrumentation in 2018.
{\sc Gerda} operated three types of enriched HPGe detectors arranged in an array of 7 strings: 7 semi-coaxial detectors, referred to as coaxial detectors for brevity, from Phase~I with a total mass of 15.6~kg; 30 Broad Energy Germanium (BEGe) detectors (20.0~kg)~\cite{Agostini:2014hra,Agostini:2019mwn}; and 5 inverted coaxial (IC) detectors (9.6~kg)~\cite{Cooper:201125,Agostini:2021wzn}, which were installed in summer 2018. 
The accumulated exposure, product of total detector mass and respective livetime, amounts to 103.7~kg$\cdot$yr for Phase II.
In order to avoid bias in the event selection criteria, {\sc Gerda} followed a strict blinding strategy, where events within a $Q_{\beta\beta}\pm25$~keV energy window were processed only after the analysis had been finalized.

In $0\nu\beta\beta$ decay, the two electrons deposit their energy in a small volume (about 1~mm$^3$ \cite{Abt:2007iy}) of a germanium detector producing single-site events (SSEs).
On the other hand, background events consisting of $\gamma$ rays from natural radioactivity interact mostly via Compton scattering producing events with multiple separated energy depositions (multi-site events, MSEs).
Therefore, events with energy depositions in multiple germanium detectors or in the LAr volume around the detector array are discarded as background events.
The time structure of the germanium signals is used to identify MSEs in a single detector and additionally recognize events close to the detector surface, together referred to as pulse shape discrimination (PSD).
In this paper, we present the PSD techniques applied to the {\sc Gerda} Phase~II data that allow for a background-free operation of the experiment in combination with the LAr veto~\cite{Agostini:2017iyd}.
The PSD methods have been improved and extended compared to Phase~I~\cite{Agostini:2013jta}. 
They were applied consistently to the complete Phase~II dataset which was split in two parts, one before and the other after the 2018 upgrade, to account for changes in the readout electronics (cables and cross-talk).
Alternative methods, e.g. applied to parts of the data in intermediate releases~\cite{Agostini:2017iyd,Agostini:2019hzm}, are summarized in the Appendix.


\section{Signal formation, readout and processing in germanium detectors}
\label{sec:pulses}

\begin{figure*}[p]
    \centering
    \includegraphics[width=0.98\textwidth]{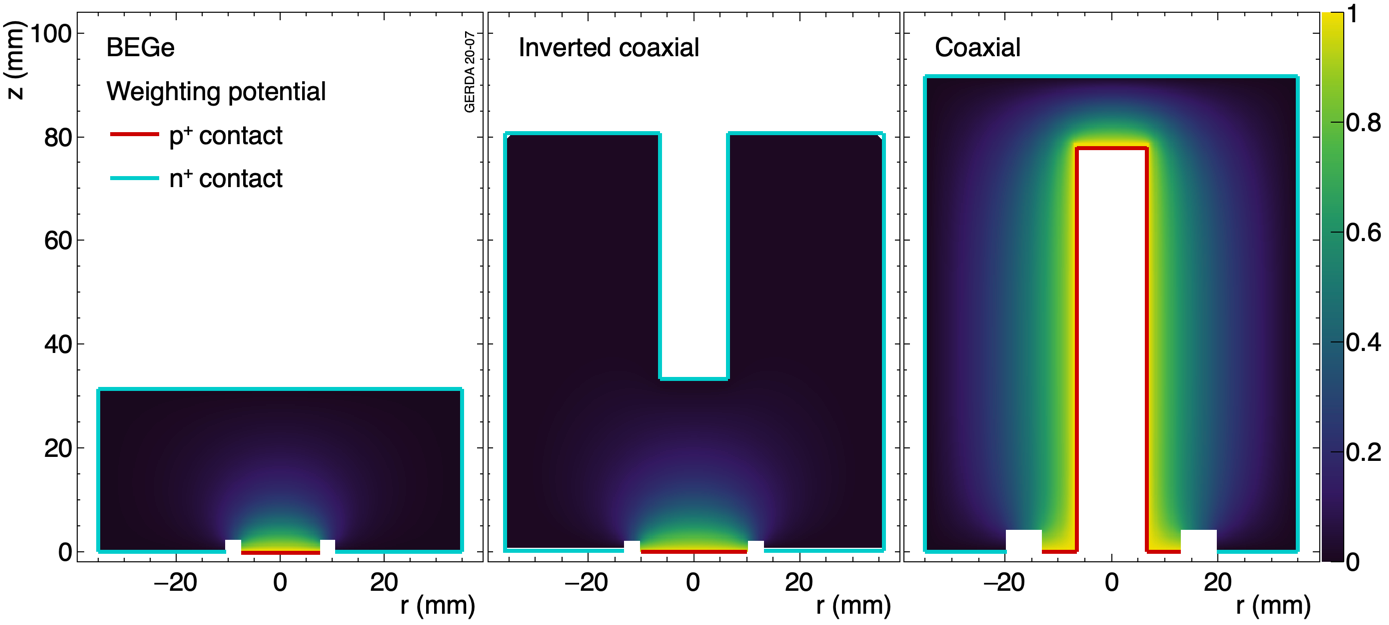}
    \caption{Geometry and weighting potential of a typical BEGe, an inverted coaxial and a coaxial detector. The p$^+$ and n$^+$ contacts are indicated schematically.}
    \label{fig:wpot}
\end{figure*}

{\sc Gerda} used p-type HPGe diodes of three different geometries called coaxial, BEGe and inverted coaxial (Fig. \ref{fig:wpot}). 
They all have a relatively thick ([0.8-2.6]~mm) n$^{+}$ electrode, formed by lithium diffusion, and a thin ($\sim$300~nm) p$^{+}$ electrode created by boron implantation. 
The p$^{+}$ and n$^{+}$ electrodes are separated by a groove with non-conducting surface. 
Because of their small p$^{+}$ electrode the BEGe and IC detectors belong to the class of point-contact detectors \cite{Luke:1989pcd} which exhibits intrinsic performance advantages with respect to energy resolution and pulse shape discrimination.

The Ge detectors are operated under reverse bias voltage such that almost the entire volume is depleted of free charge carriers. 
An interaction in the active volume creates a number of electron-hole pairs proportional to the
deposited energy.  
The charge carriers drift according to the electric field created by both the positive bias voltage applied to the n$^{+}$ contact and the volume charge density due to the net bulk impurity concentration. 
The electrons are collected at the n$^{+}$ contact, the holes at the p$^{+}$ contact which is used for readout.
The n$^{+}$ layer covers most of the crystal surface and features a large Li concentration.
Exhibiting zero electric field beyond the p-n junction till the outer surface, a charge created in this layer will only experience thermal diffusion with two possible outcomes: recombine (loss) or reach the depleted volume (collection). As a result, two generic regions can be identified, the dead layer with high probability of no charge collection and a transition layer with partial charge collection.

During the drift of charge carriers, charge is induced on the readout contact as described by the Shockley-Ramo theorem~\cite{HE2001250}:
\begin{equation}
    Q(t) = -Q_0 \left[ \Phi^w(\mathbf{r}_h(t)) - \Phi^w(\mathbf{r}_e(t)) \right]
\end{equation}
where $Q_0$ is the total charge carried by the holes or electrons and $\Phi^w(\mathbf{r}_{h/e}(t))$ is the weighting potential along the drift path of holes or electrons.
The weighting potential is shown in Fig.~\ref{fig:wpot} for the three different detector geometries used in {\sc Gerda} also indicating the geometry of the p$^+$ and n$^+$ surfaces.
The weighting potentials have been calculated using the AGATA Detector Library~\cite{Bruyneel:2016zih} pulse shape simulation package.

\begin{figure*}[p]
    \centering
    \includegraphics[width=0.98\textwidth]{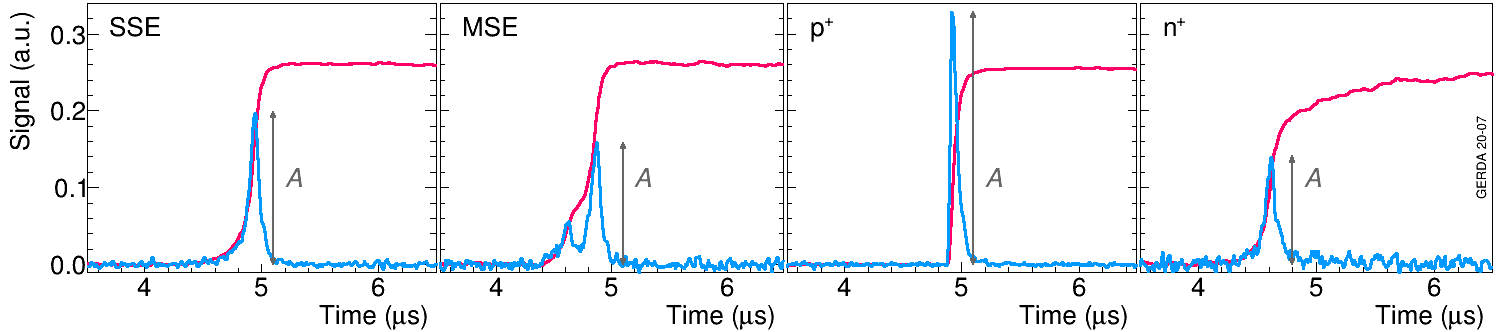}
    \caption{Examples of different type of normalized charge pulses (red) along with the derived current pulses (blue) from a BEGe detector: single-site event, multi-site event, event near the p$^+$ contact and event near the n$^+$ contact with incomplete charge collection.}
    \label{fig:bege_signals}
\end{figure*}

Due to their small p$^+$ contact, BEGe and IC detectors have a weighting potential distribution that is very small in most of the volume and sizable only close to the p$^+$ contact. 
This results in similar waveforms $Q(t)$ from interactions in a large part of the volume.
Multiple energy deposits can be treated as a superposition of single interactions.
In Fig.~\ref{fig:bege_signals}, normalized example pulses from a BEGe detector are shown for a SSE, a MSE, an event close to the p$^+$ contact and an event near the dead layer of the n$^+$ contact with incomplete charge collection. 
As shown, surface events produce characteristic pulse shapes being fast close to the p$^+$ contact due to the strong electric field and slow near the n$^+$ contact due to the weak electric field and the transition layer.

\begin{figure*}[p]
    \centering
    \includegraphics[width=0.98\textwidth]{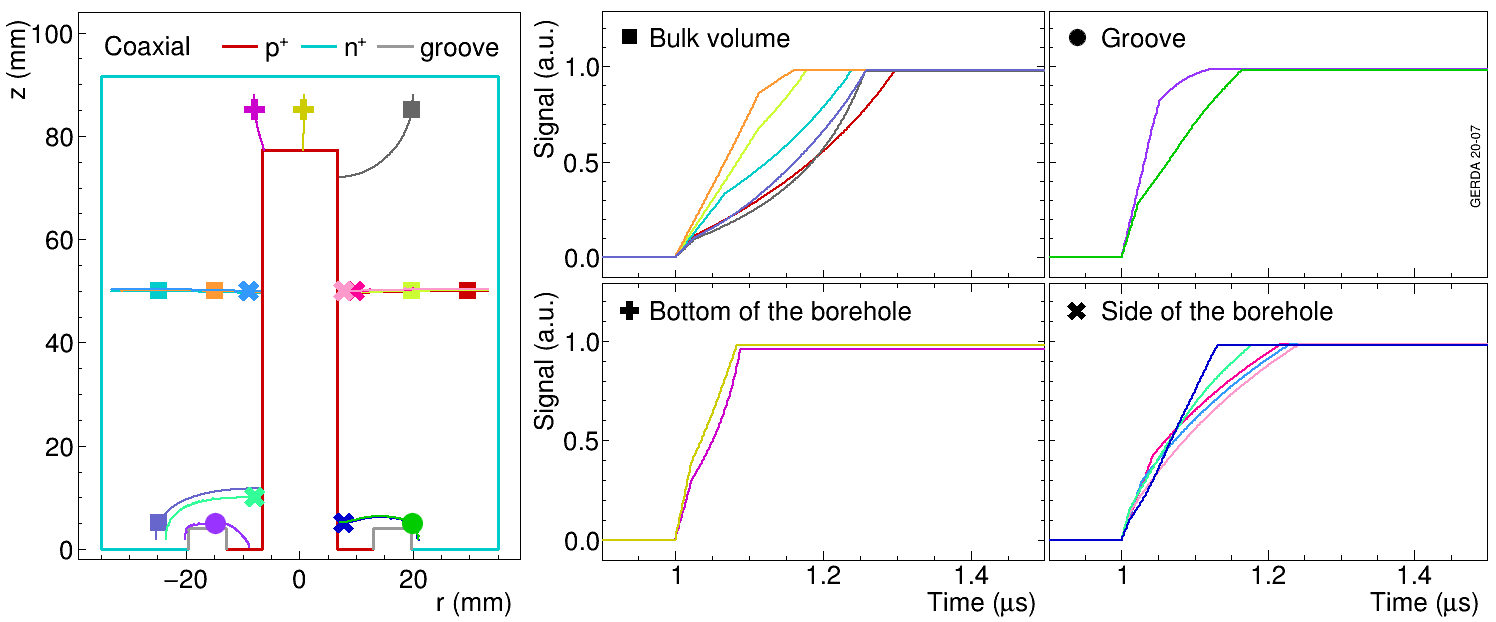}
    \caption{Simulated charge pulses in indicated parts (bulk, groove, bottom/side of
        borehole) of a typical coaxial detector.
        The electronic response of the readout chain and noise are not included here.}
    \label{fig:coax_signals}
\end{figure*}

In coaxial detectors, both the hole and electron drift play a role in the pulse formation, which result in different pulse shapes throughout the volume of the detector.
In Fig.~\ref{fig:coax_signals}, simulated example pulses from different parts of the detector are shown.
Similarly to BEGe detectors, coaxial detectors also show special pulse shape characteristics in case of surface events. Indeed, energy deposits near the groove or the bottom of the borehole cause faster pulses.

The signals induced on the p$^+$ contact of the {\sc Gerda} detectors are read out by charge sensitive amplifiers located in the LAr about 35~cm above the detector array.
The signals are digitized at 25~MHz for $160~\mu $s and at 100~MHz for 10~$\mu $s.
Both traces are centered at the rising edge of the charge pulse $Q(t)$.
The offline analysis of the digitized signals follows the procedures described in~\cite{Agostini:2011mh,Agostini:2011xe}.
The 25~MHz traces used for the energy reconstruction ensure the excellent energy resolution, while the 100~MHz traces are used in the PSD methods presented in the following sections.
The energy estimator $E$ is reconstructed with a zero-area cusp filter~\cite{Agostini:2015pta}, whose parameters are optimized for each detector and calibration run.


\section{Overview of event samples and discrimination methods}
\label{sec:events}

\begin{figure}[htb]
    \centering
    \includegraphics[width=0.48\textwidth]{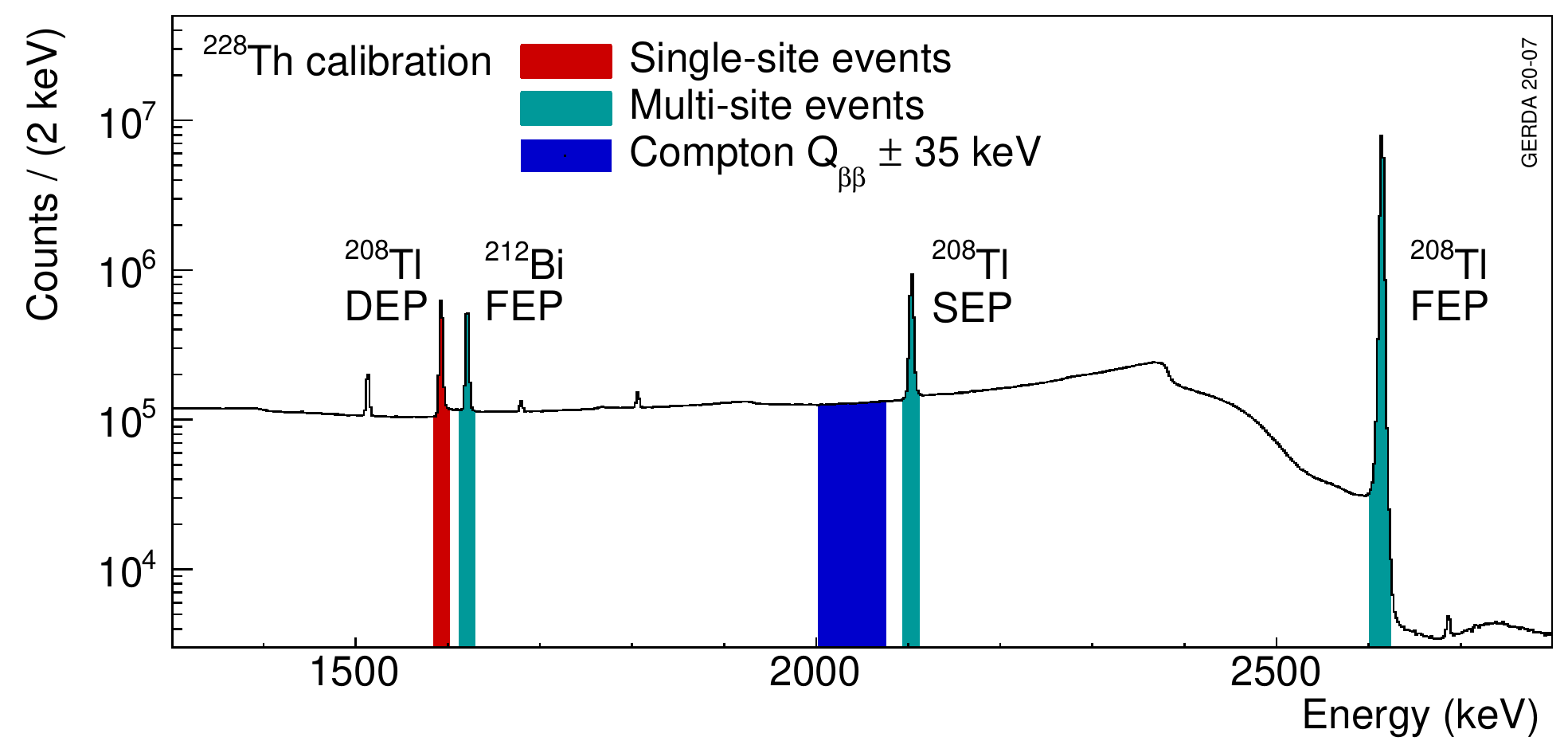}
    \caption{Calibration spectrum highlighting the different event samples used in the pulse shape analysis.}
    \label{fig:calspectrum}
\end{figure}

Weekly calibration runs with $^{228}$Th sources are performed to determine the energy scale and resolution of the detectors \cite{Agostini:2021xyz} and to calibrate and train the PSD techniques.
Figure~\ref{fig:calspectrum} shows a calibration spectrum highlighting the different event samples used in pulse shape analysis.
The most prominent feature is the full energy peak (FEP) at 2615~keV from $^{208}$Tl decay. 
Its double escape peak (DEP) at 1593~keV is used as a sample of SSEs as the electron and positron from pair production deposit their energy in a small volume and both annihilation $\gamma$ rays leave the detector.
The FEP at 1621~keV from $^{212}$Bi is used as a sample of MSEs that is sufficiently near in energy to the DEP in order to avoid noise dependent biases.
To test the performance of MSE rejection, the FEP and single escape peak (SEP) of $^{208}$Tl, mostly featuring MSEs are used while the Compton continuum region around $Q_{\beta\beta}\pm35$~keV (CC($Q_{\beta\beta}$)) serves to estimate the background rejection in the $0\nu\beta\beta$ decay signal region.

In physics data, the standard neutrino accompanied double-$\beta$ ($2\nu\beta\beta$) decay provides another sample of signal-like events that is equally distributed in the whole detector volume and used for the investigations in the following.
After applying the LAr veto cut \cite{Agostini:2017hit}, about 97\% of the events in the $1000-1300$~keV region originate from $2\nu\beta\beta$ decays.
Beside MSEs from $\gamma$ rays, the physics data have a significant amount of surface events from $\alpha$ and $\beta$ decays that can be discriminated thanks to their specific pulse shape.
The physics spectrum at low energies is dominated by $\beta$ decays of $^{39}$Ar up to its $Q$-value of 565~keV.
However, these events are not used in the pulse shape analysis due to their relatively high noise.
A prominent background source at $Q_{\beta\beta}$ is the $\beta$ decay of $^{42}$K, which is produced as a progeny of the long-lived $^{42}$Ar and has a $Q$-value of 3525~keV.
Beta particles deposit their energy in germanium within a few mm resulting in events partly in the dead and transition layers of the n$^+$ surface.
Such n$^+$ surface events can induce slow pulse shapes with incomplete charge collection.
Apart from possible HPGe bulk contamination, that have been shown to be insignificant \cite{Abramov:2019hhh,Agostini:2017lim}, $\alpha$ particles can only reach the active volume of the detector at the thin p$^+$ contact or at the non-conducting groove producing pulse shapes with fast rise as shown in Figs.~\ref{fig:bege_signals} and \ref{fig:coax_signals}.
A clean sample of $\alpha$ surface events is found in physics data above the $^{42}$K $Q$-value.
The most prominent structure at these energies is a broad peak at 5304~keV, the $\alpha$ energy of the $^{210}$Po $\alpha$ decay ($^{238}$U decay chain) \cite{Abramov:2019hhh,Agostini:2014}.

Due to their different geometries, BEGe and IC detectors are treated separately from coaxial detectors in the pulse shape analysis.
In  the  case  of the BEGe and IC detectors one parameter, $A/E$, is used to classify background events, where $A$ is the maximum current amplitude as indicated in Fig.~\ref{fig:bege_signals} and $E$ is the energy. 
As MSEs and surface events at the n$^+$ contact are characterized by longer, i.e. wider current pulses, they feature a lower $A/E$ value compared to SSEs, while surface events at the p$^+$ contact show a higher $A/E$ value~\cite{Budjas:2009zu}.
Therefore, rejecting events on both sides of the $A/E$ distribution of SSEs enhances the signal to background ratio.
The details of the $A/E$ analysis are presented in Sec.~\ref{sec:bege}.
In the case of coaxial detectors an artificial neural network (ANN \cite{TMVA}) is used to discriminate SSEs from MSEs similar to the approach applied in Phase~I~\cite{Agostini:2013jta}.
To discard events close to the p$^+$ contact, a dedicated cut on the risetime of the pulses is applied.
The training and optimisation of the ANN and risetime cuts for coaxial detectors is described in Sec.~\ref{sec:coax}.
An additional cut is applied to all detectors to remove events with slow or incomplete charge collection. 
These events arise from energy depositions in a non-depleted volume (n$^+$ layer or insulating groove).
These events are identified through the difference between two energy estimates performed using the same digital filter but different shaping times as summarized in Sec.~\ref{sec:deltaE}.
The signal efficiency of $0\nu\beta\beta$ decay is estimated using the survival fraction of DEP and $2\nu\beta\beta$ decay events by taking into account the energy dependence of the different PSD techniques.
The details of this extrapolation to $Q_{\beta\beta}$ including systematic uncertainties are found in Sec.~\ref{sec:efficiency} and \ref{app:pss}.


\section{The A/E method for BEGe and IC detectors}
\label{sec:bege}

The amplitude of the current pulse $A$ is computed after applying 3 times a moving window average (MWA) filter with 50~ns length and interpolating the pulse down to 1~ns sampling time.
This filtering procedure optimizes the high frequency noise attenuation while preserving the pulse shape information.
The energy estimator $E$ is determined by a pseudo Gaussian filter with a shaping time of 10~$\mu$s.
$A$ is then divided by $E$ before calibration, providing the raw $A/E$ for each pulse.
The raw $A/E$ is then corrected for time stability and energy dependence before a cut value is defined.

For each calibration run, the $A/E$ distribution of events in the $1000-1300$~keV region is fitted with a Gaussian (SSEs) and a low-side tail (MSEs) as described in~\cite{Agostini:2013jta}.
The position of the Gaussian $\mu_{A/E}$ from each calibration of the four years of data taking is used to define stable time periods where the raw $A/E$ changes by less than its $\sigma_{A/E}$ resolution.
Instabilities are mostly related to hardware changes and a few detectors show a small systematic drift of the raw $A/E$.
Physics data between the stable periods are removed from the analysis causing a few percent exposure loss.
After normalizing the raw $A/E$ by the average $\mu_{A/E}$ within a given time period, the data of all calibrations are merged, only separating before and after the upgrade of 2018.

\begin{figure}[t]
    \centering
    \includegraphics[width=0.48\textwidth]{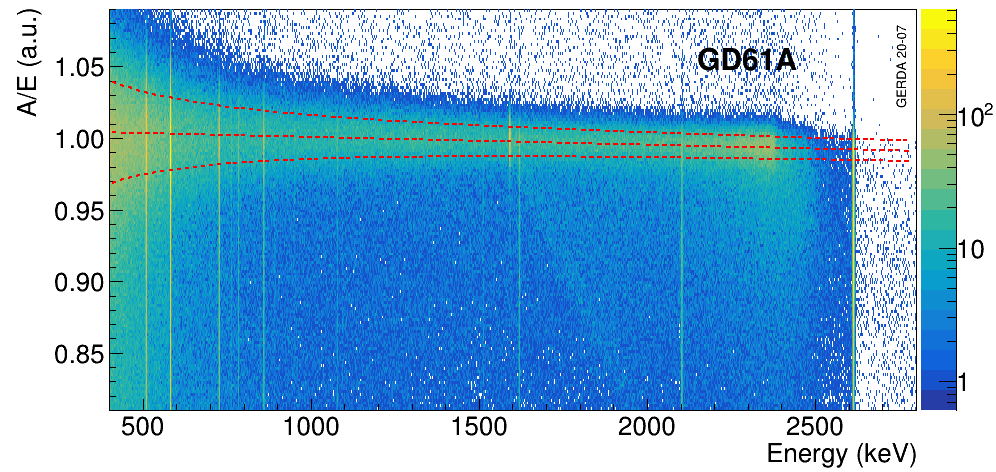}
    \includegraphics[width=0.48\textwidth]{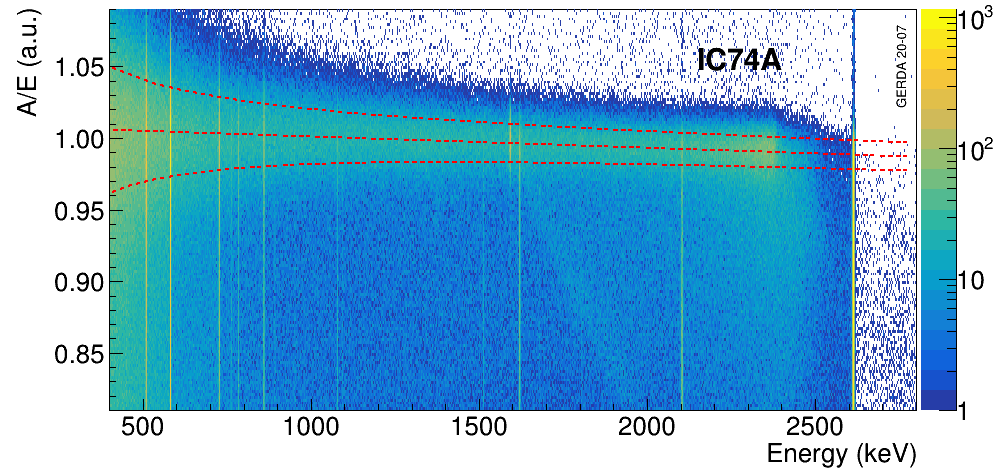}
    \caption{Distribution of $A/E$ as a function of energy from $^{228}$Th calibration data for the BEGe detector GD61A (top) and the IC detector IC74A (bottom). The dashed lines indicate the linear energy dependence and the 1-$\sigma$ width of SSEs.}
    \label{fig:aoe_energy}
\end{figure}

The $A/E$ of SSEs in the Compton continuum of the merged calibration data show a small linear energy dependence of a few percent per MeV, due to the larger charge cloud size at higher energies that broadens the current pulse.
The energy dependence of $\mu_{A/E}(E)$ and $\sigma_{A/E}(E)$ is described by a linear and a $\sqrt{b+c/E^2}$ type of function, respectively, as shown in Fig.~\ref{fig:aoe_energy} for a BEGe (GD61A) and an inverted coaxial (IC74A) detector as examples.
In addition to the correction for the energy dependence, $A/E$ is normalized to the mean of the $A/E$ distribution of the DEP, which lies about 0.25\% above the SSE band.

The cut values for each detector are determined on the energy-independent $A/E$ classifier defined as $\zeta=([A/E]\,/\,\mu_{A/E}(E)-1)\,/\,\sigma_{A/E}(E)$.
Its distribution is centered around zero and has a standard deviation of one for SSEs.
The low-side $A/E$ cut against MSEs and n$^+$ surface events is chosen to yield a DEP survival fraction of 90\%.
The resulting cut values range from $\zeta=-1.9$ to $-1.2$ and from $\zeta=-1.9$ to $-1.7$ for BEGe and IC detectors, respectively.
The high-side $A/E$ cut against p$^+$ surface events is chosen at $\zeta=3.0$ for each detector, in order to reject all $\alpha$ events in physics data above 3525~keV.
It has been shown that the high-side $A/E$ cut discards events, including degraded $\alpha$ events, from a small volume around the p$^+$ contact \cite{Comellato:2020ljj} causing the survival fraction of events after the high-side $A/E$ cut to be proportional to the detector mass.

\begin{figure}[t]
    \centering
    \includegraphics[width=0.48\textwidth]{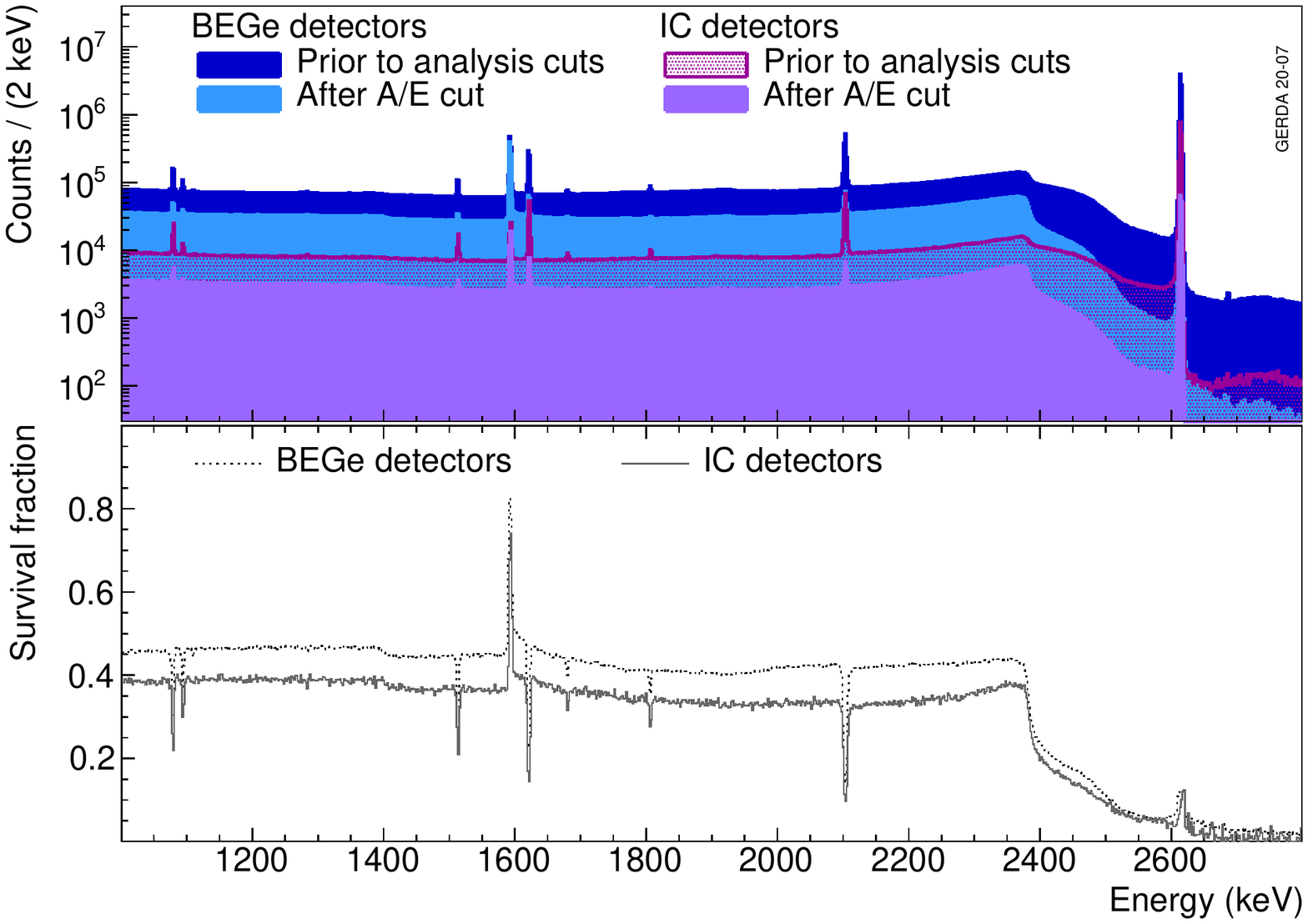}
    \caption{Energy distribution of events from calibration data before and after the $A/E$ cut (top) and their ratio (bottom) for BEGe and IC detectors.}
    \label{fig:aoe_calib}
\end{figure}

Fig.~\ref{fig:aoe_calib} shows for BEGe and IC detectors the energy distribution of calibration data before and after the $A/E$ cuts described above as well the corresponding survival fractions.
As expected the survival fraction of events in the FEPs and SEP is much lower than in the DEP.
Events in the Compton continuum are discarded with about the same probability independent of their energy but depending on the overall detector size, and more generally speaking from the detector type.
IC detectors discard a higher fraction of events because of the higher probability of multiple scattering of $\gamma$ rays due to their larger size.

\begin{figure}[b]
    \centering
    \includegraphics[width=0.48\textwidth]{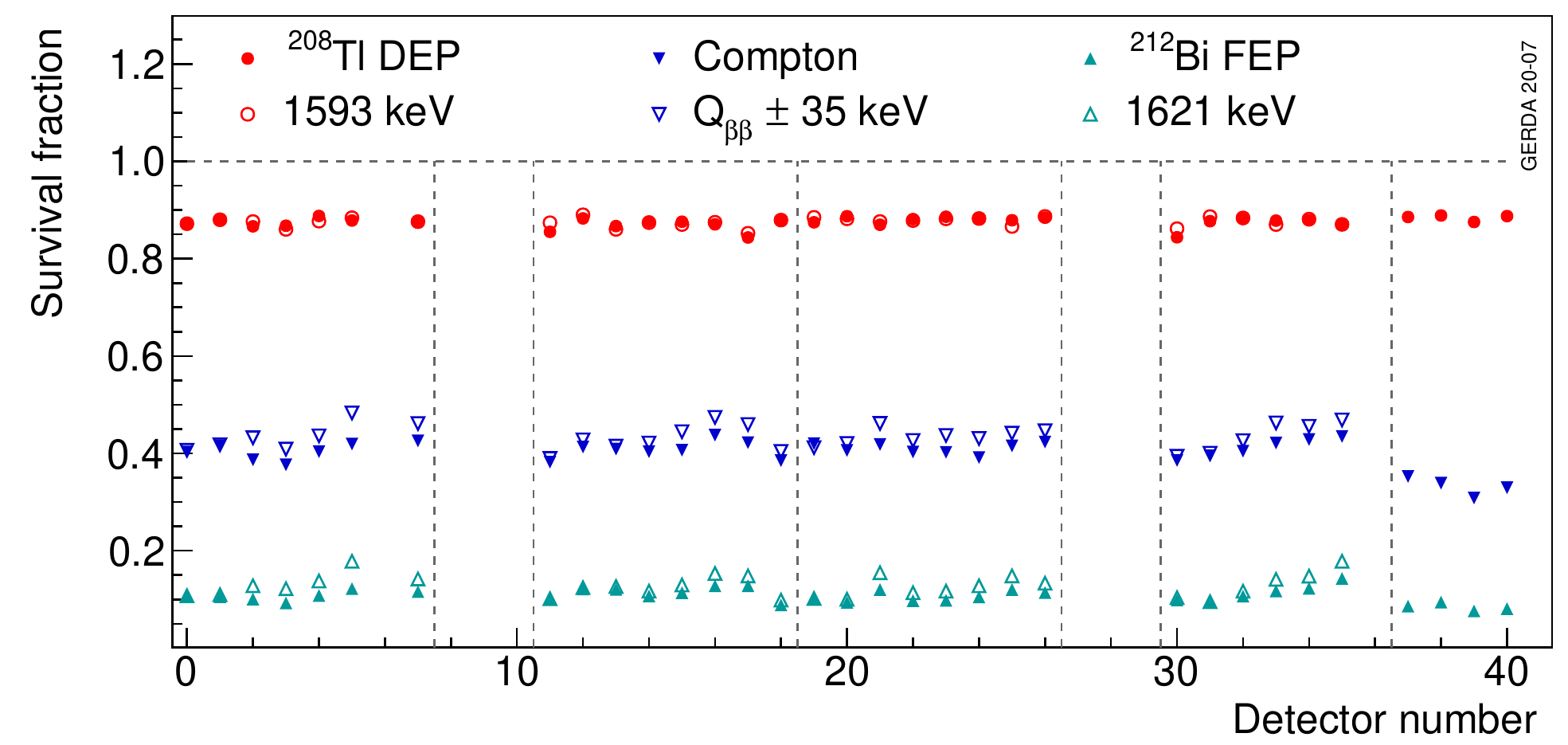}
    \caption{Survival fraction of events in the $^{208}$Tl~DEP, $^{212}$Bi~FEP and CC($Q_{\beta\beta}$) for each detector (see Table \ref{tab:det-eff} for detector numbers and types). Open (filled) symbols show the calibration dataset before (after) the upgrade. The dashed lines separate the detector strings in the array. The uncertainties are only statistical and smaller than the markers.}
    \label{fig:aoe_eff}
\end{figure}

The survival fractions of events in the DEP, FEP and CC($Q_{\beta\beta}$) are shown in Fig.~\ref{fig:aoe_eff} for each detector before and after the upgrade.
The DEP survival fractions are slightly smaller than 90\% due to the high $A/E$ cut.
The rejection of MSEs shows a small dependence on the detector position in the string because of different electronic noise conditions. 
This effect was reduced after the upgrade.
The IC detectors (detector numbers above 35) reject MSEs more efficiently than BEGe detectors.

\begin{figure}[t]
    \centering
    \includegraphics[width=0.48\textwidth]{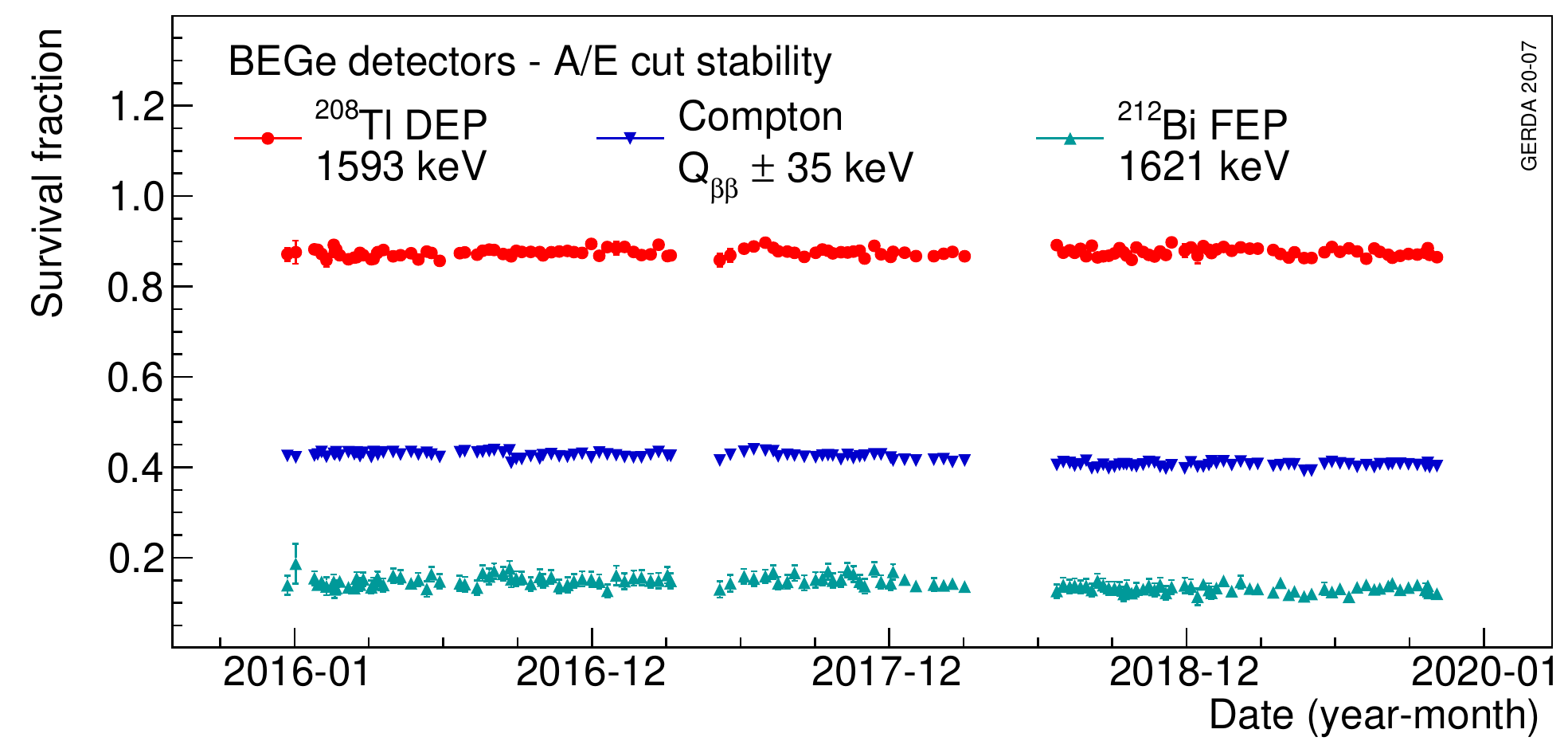}
    \caption{Average survival fractions of events in the $^{208}$Tl~DEP, $^{212}$Bi~FEP and CC($Q_{\beta\beta}$) for BEGe detectors as a function of time. Each data point represents a calibration run with its statistical uncertainty.}
    \label{fig:aoe_stability}
\end{figure}

In order to check the validity of the $A/E$ corrections and cuts in the whole dataset, survival fractions of the usual event samples are studied for each calibration run.
The average survival fraction from all BEGe detectors is shown for each calibration in Fig.~\ref{fig:aoe_stability} as a function of time.
The $A/E$ cut shows, for both BEGe and IC detector types, a stable behaviour at the 3\% relative level during the whole data collection period when applied to Compton continuum events at $Q_{\beta\beta}$.
Residual instabilities stem on one side from changes in the detector gain or leakage current and on the other side from statistical fluctuations.

\begin{figure}[b]
    \centering
    \includegraphics[width=0.48\textwidth]{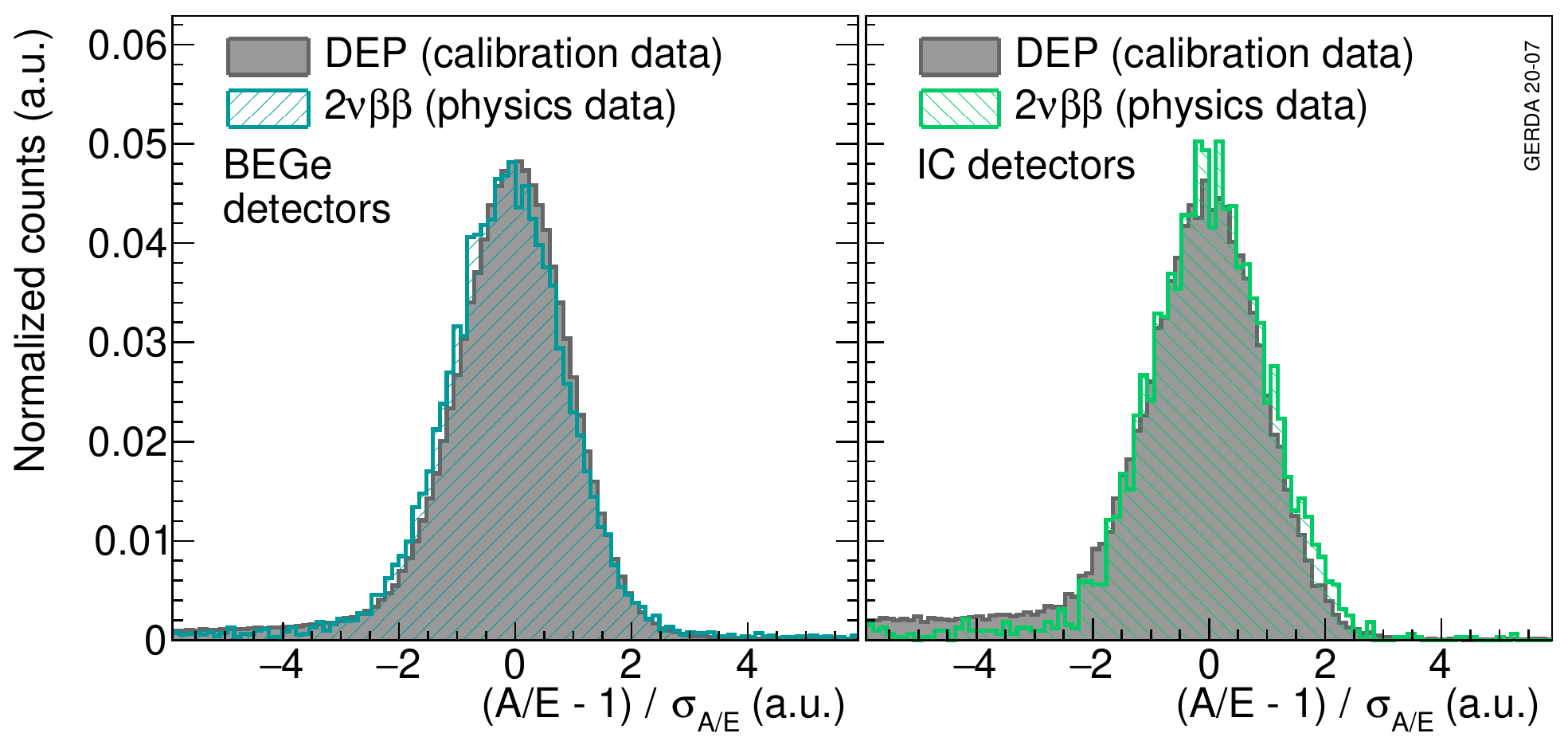}
    \caption{$A/E$ classifier distributions after LAr veto comparing events in the DEP from calibration data and $2\nu\beta\beta$ decay events from physics data from BEGe (left) and IC (right) detectors. The histograms are normalized to their integrals.}
    \label{fig:aoe_compare}
\end{figure}

The raw $A/E$ is corrected in the same way for physics data and the same cut values are applied as in calibration data.
In order to check the validity of the corrections and the cut values, Fig.~\ref{fig:aoe_compare} shows a comparison of the $A/E$ classifier between DEP events from calibration data and $2\nu\beta\beta$ events from physics data for BEGe and IC detectors.
By construction, the $A/E$ classifier peaks at 0 and has a width of about 1 for these SSEs.
The agreement between physics and calibration data is satisfactory and confirms the applied correction procedure.
For each detector, the residual difference of the average $A/E$, between calibration and physics data, is included in the systematic uncertainties by shifting the cut value accordingly.

\begin{figure}[t]
    \centering
    \includegraphics[width=0.48\textwidth]{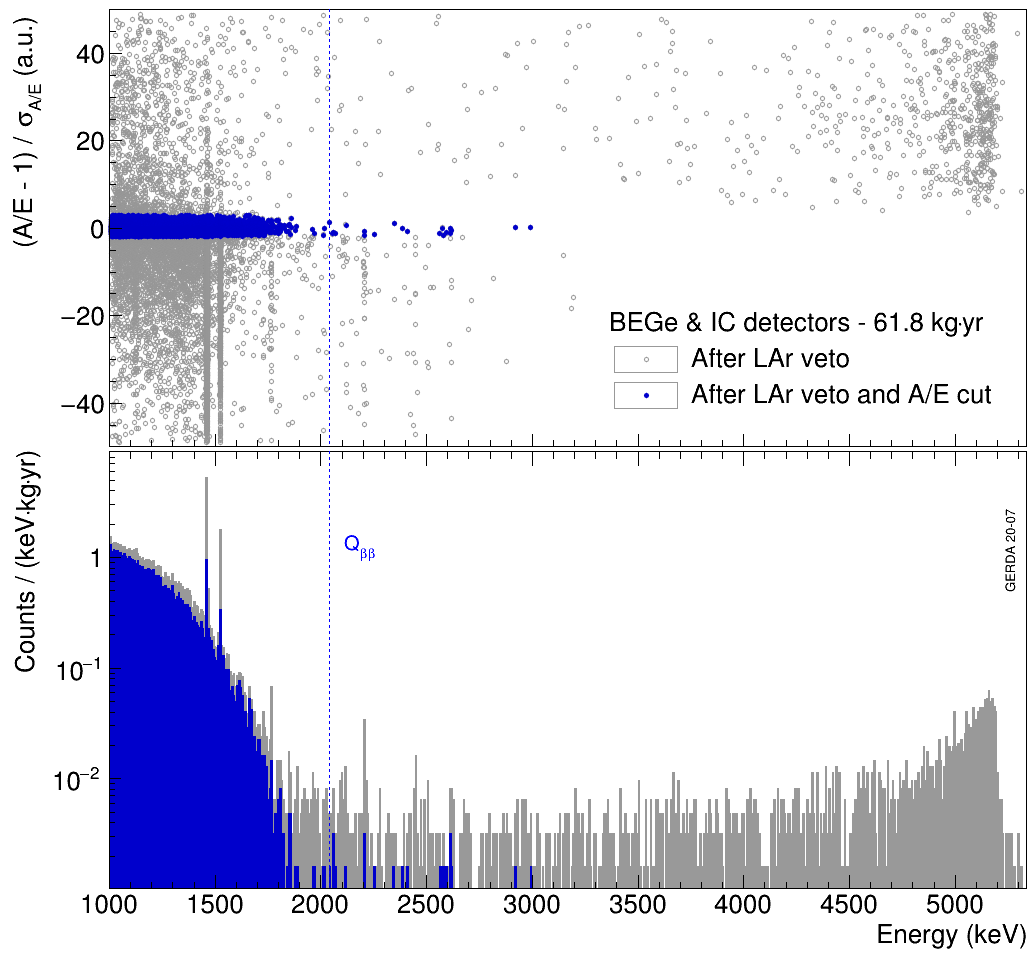}
    \caption{Energy distributions of all {\sc Gerda} Phase II physics data from the BEGe and IC detectors before (grey) and after (blue) the $A/E$ cut.
    The position of $Q_{\beta\beta}$ is indicated.
    The prominent $^{40}$K and $^{42}$K lines are located at 1461 keV and 1525 keV, respectively.}
    \label{fig:aoe_phys}
\end{figure}

The energy distributions of events before and after the $A/E$ cut are shown in Fig.~\ref{fig:aoe_phys} for the whole physics data corresponding to 61.8~kg$\cdot$yr exposure from BEGe (53.3~kg$\cdot$yr) and IC (8.6~kg$\cdot$yr) detectors.
Events in the $2\nu\beta\beta$ decay region survive the cut with a high probability, while the $^{40}$K and $^{42}$K peaks at 1461 keV and 1525 keV, respectively, mostly featuring MSEs, are reduced significantly.
High energy events above 3525~keV coming from p$^+$ surface events are all discarded by the high $A/E$ cut by definition.
In a 240 keV wide window around $Q_{\beta\beta}$, only 7 events in BEGe detectors and 1 event in IC detectors survive the LAr veto and $A/E$ cuts.
This results in the corresponding unique background indices\footnote{The background index is evaluated in the range between 1930 and 2190~keV without the two intervals ($2104\pm5$) and ($2119\pm5$)~keV due to known $\gamma$ rays and without the signal interval ($Q_{\beta\beta}\pm5$) keV~\cite{Agostini:2020prl}. The quoted uncertainties are statistical only.} of {\sc Gerda} in the region of interest of $5.5_{-1.8}^{+2.4} \cdot 10^{-4}$~counts/(keV$\cdot$kg$\cdot$yr) and $4.9_{-3.4}^{+7.3} \cdot 10^{-4}$~counts/(keV$\cdot$kg$\cdot$yr) for BEGe and IC detectors, respectively.

\section{The ANN and risetime methods for coaxial detectors}
\label{sec:coax}

The 100~MHz waveform traces are used to compute the artificial neural network input variables (IVs) and the risetime.
They are first filtered with a MWA of 30~ns width three times.
They are subsequently baseline subtracted and normalized between 0 and 1 by the amplitude of the 25~MHz pole-zero corrected traces.
This amplitude is provided by a trapezoidal filter with a typical precision of 0.2\% at 2~MeV.

\begin{figure}[htb]
    \centering
    \includegraphics[width=0.48\textwidth]{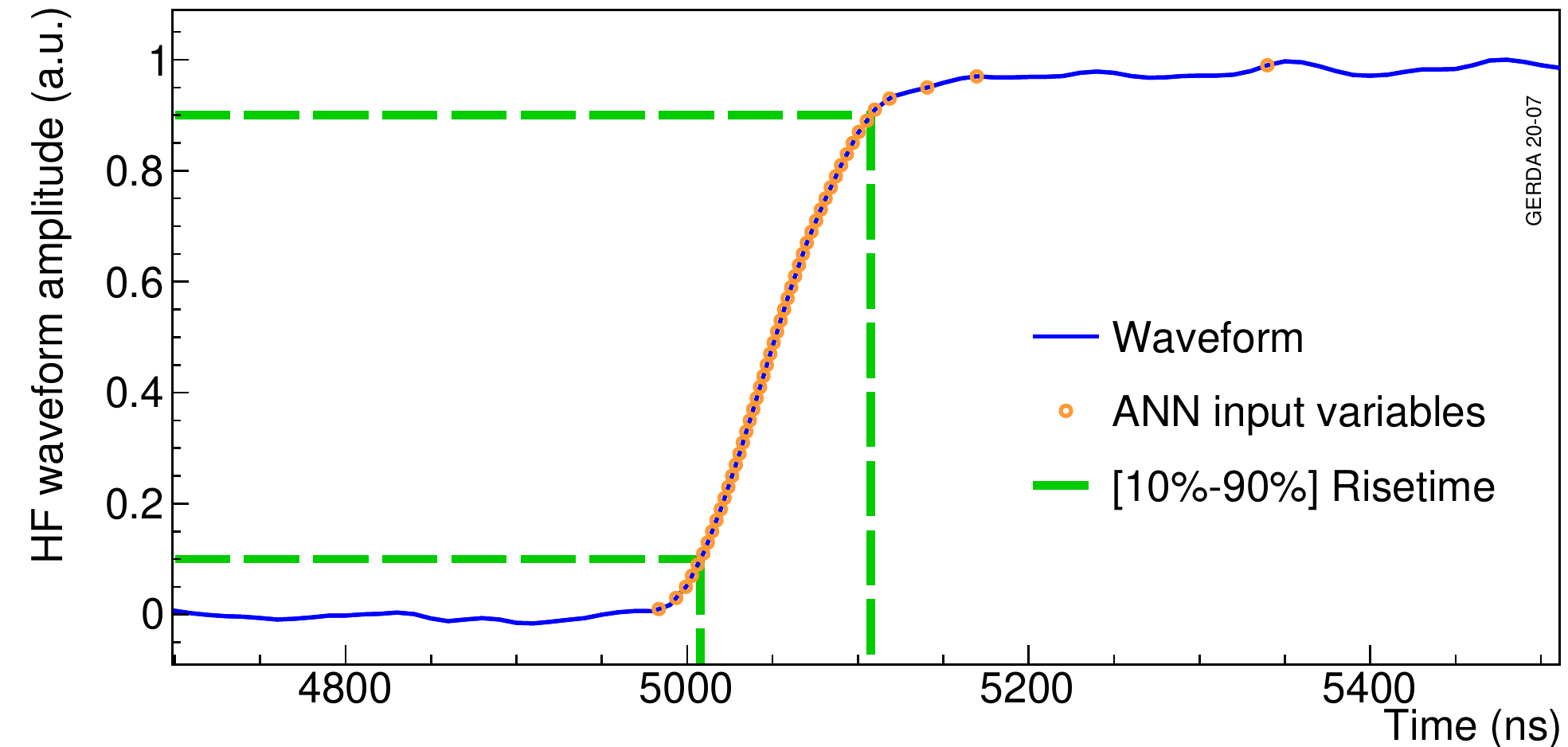}
    \caption{Zoom on a typical normalized 100~MHz trace of a coaxial detector.
    The 50 ANN input variables (red circles) and rise time estimates (dashed green) are indicated.}
    \label{fig:ANN_RT_pulse}
\end{figure}

The ANN IVs are a list of the 50 times at which the resulting waveform reaches [1\%, 3\%, ..., 99\%] of its amplitude (see Fig.~\ref{fig:ANN_RT_pulse}).
An interpolation of the 10~ns wide gaps between data points allows for a more precise estimation.
These IVs are computed for all physics and calibration events found in {\sc Gerda}. 
However, given the degraded signal-to-noise ratio at low energy, only events above 1000~keV are considered in this pulse shape analysis to avoid loosely constrained energy dependence correction.
Calibration runs are used to optimize the discrimination of SSEs from MSEs. 
The network is built on two hidden layers with 50 and 51 neurons, using the {\sc TMVA - MLPBFGS} algorithm \cite{TMVA}.

\begin{figure}[htb]
    \centering
    \includegraphics[width=0.48\textwidth]{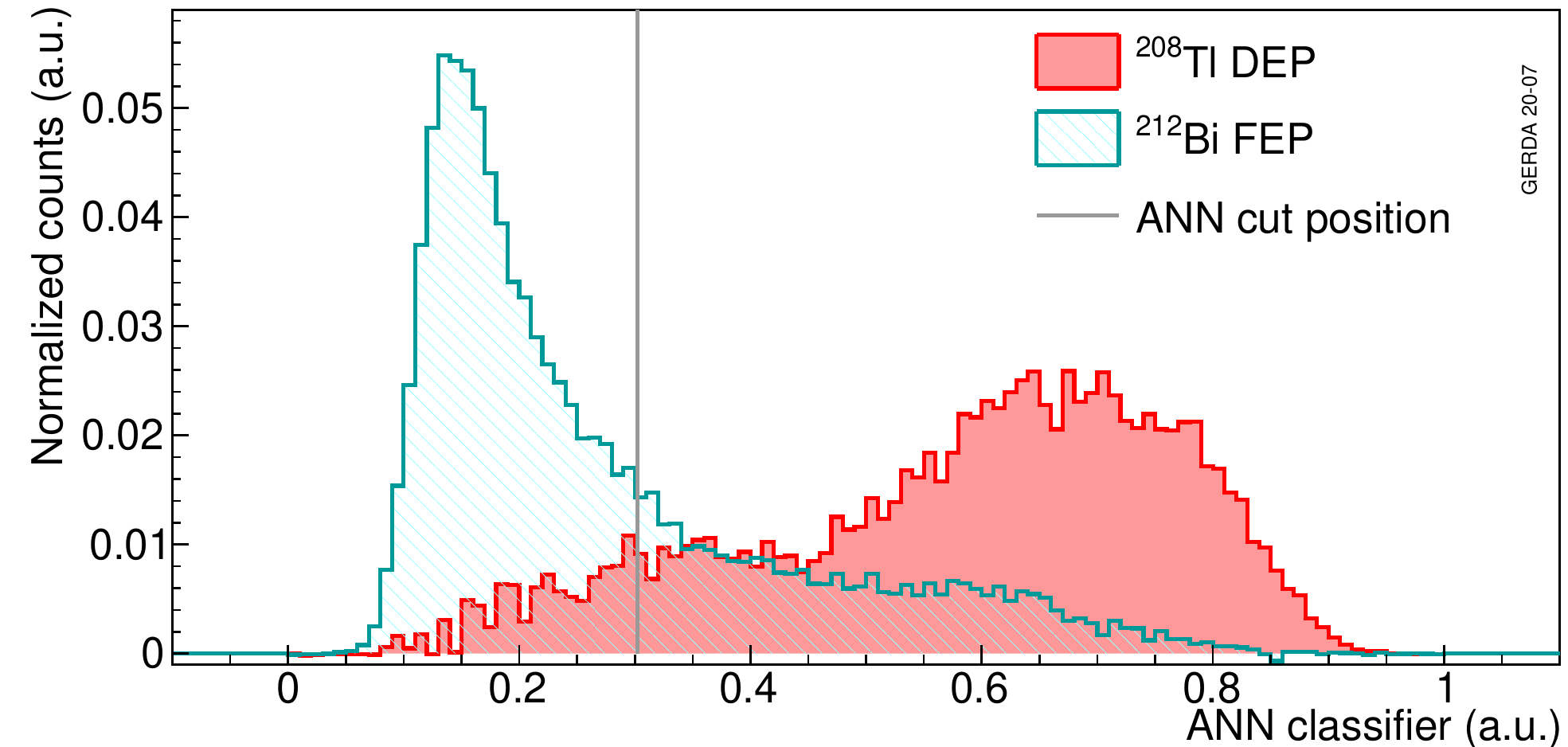}
    \caption{Trained ANN classifier values for events in the $^{208}$Tl~DEP (SSEs) and $^{212}$Bi~FEP (MSEs) from the ANG5 detector. The histograms are normalized to their integrals.}
    \label{fig:ANN_classification}
\end{figure}

Fig.~\ref{fig:ANN_classification} shows as an example the classifier distributions from the ANN training of the ANG5 detector with the events from the indicated DEP and FEP peaks. The distributions from the Compton events under the peaks are statistically subtracted using the distributions of the events in the energy side-bands of the peaks.
The lower and upper side-bands are defined by selecting events falling in the  [$-9 \sigma$,$-4.5 \sigma$] and [$4.5 \sigma$,$9 \sigma$] energy regions where $2.355 \cdot \sigma$ is the full width at half maximum used to quantify the energy resolution of the Ge detectors. The indicated ANN cut keeps 90\% of the events in the DEP peak.

Due to the significant change in hardware, the data taken before and after the 2018 upgrade periods have been trained separately.
Finer splittings of the data have yielded signal efficiencies and background rejection values which agreed on the one-percent level.
As a result of the limited statistics, a minimal number of two datasets was preferred.
Typically, about 10000 and 15000 events enter the signal (DEP) and background ($^{212}$Bi FEP) samples, respectively, for the ANN training of each detector.
Similarly to the $A/E$ method, the ANN cut for each detector and each training period is set on the $^{208}$Tl~DEP classifier distribution such that 90\% of the events survive.

\begin{figure}[htb]
    \centering
    \includegraphics[width=0.48\textwidth]{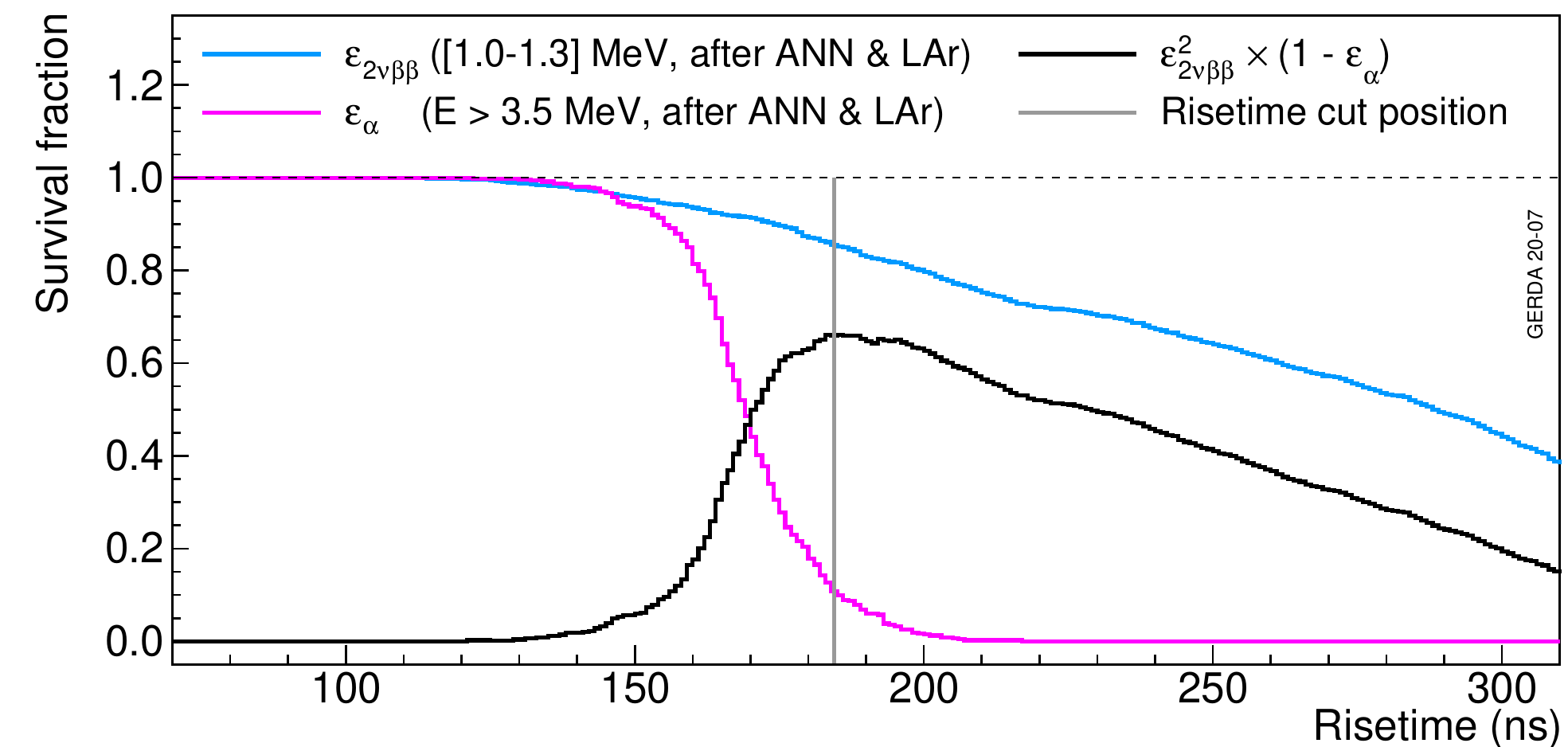}
    \caption{Survival fractions after the rise time cut for $2\nu\beta\beta$ decay and $\alpha$ particles events from the Phase II data before the upgrade of the ANG4 detector. 
    Also shown are the figure of merit and the chosen cut value.}
    \label{fig:RT_cut}
\end{figure}

The risetime is estimated after interpolating the waveform with a 1~ns time step.
From studies on the rejection of $\alpha$ particles~\cite{Lazzaro:2019pda}, the [10\%-90\%] amplitude signal risetime was selected (see Fig.~\ref{fig:ANN_RT_pulse}).
This parameter is used to reject $\alpha$ events on the p$^+$ contact that develop faster signals (see Fig.~\ref{fig:coax_signals}).
The cut definition relies on the maximization of the following figure of merit:
\begin{equation}
    f(x) = \varepsilon^2_{2\nu\beta\beta}(x) \cdot (1-\varepsilon_{\alpha}(x)),
\end{equation}
where $\varepsilon_{2\nu\beta\beta}(x)$ is the $2\nu\beta\beta$ survival fraction at risetime cut $x$ and $\varepsilon_{\alpha}(x)$ is the corresponding survival fraction of $\alpha$ events.
Only physics data after ANN-MSE and LAr veto, to increase purity of samples, are used for this figure of merit that allows to reject most of the $\alpha$ particles while preserving a high $2\nu\beta\beta$ decay signal survival fraction.
Figure \ref{fig:RT_cut} depicts an example of such a cut definition.
On average, about 90\% of the high energy $\alpha$ particles are rejected.

\begin{figure}[htb]
    \centering
    \includegraphics[width=0.48\textwidth]{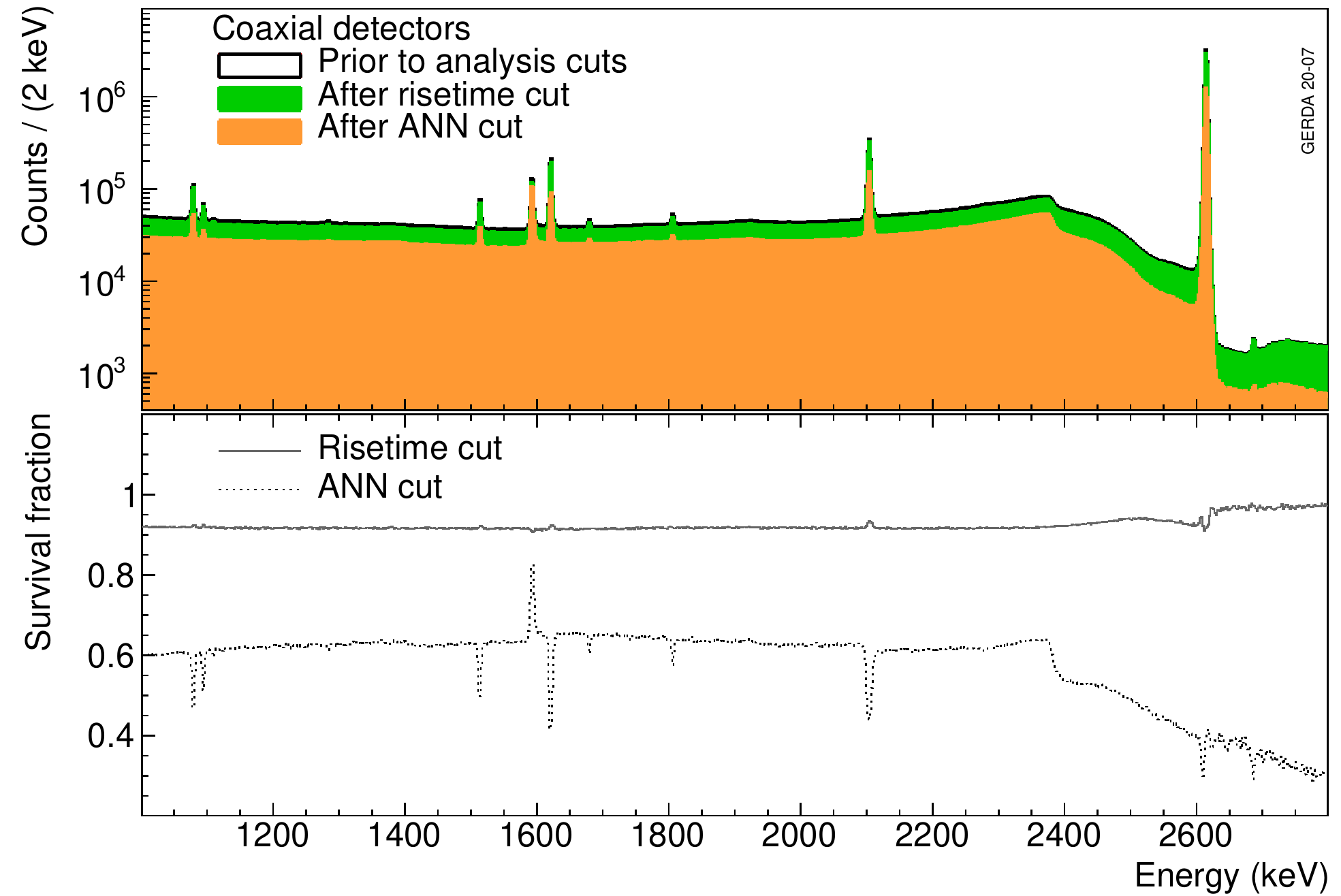}
    \caption{Energy distributions of calibration data events from the coaxial detectors before and after the ANN and risetime cuts (top) and the corresponding survival fractions (bottom).}
    \label{fig:coax_calib}
\end{figure}

The calibration energy spectrum of all coaxial detectors before and after applying the PSD cuts is shown in Fig.~\ref{fig:coax_calib}. 
As targeted, the ANN cut removes preferentially the regions highly populated by MSEs (FEPs and SEP in particular) and preserves 90\% of the $^{208}$Tl~DEP.
On the contrary, the risetime cut deployed to reject events with fast risetime is insensitive to these types of events and hence has a high survival fraction for both SSEs and MSEs.

\begin{figure}[htb]
    \centering
    \includegraphics[width=0.48\textwidth]{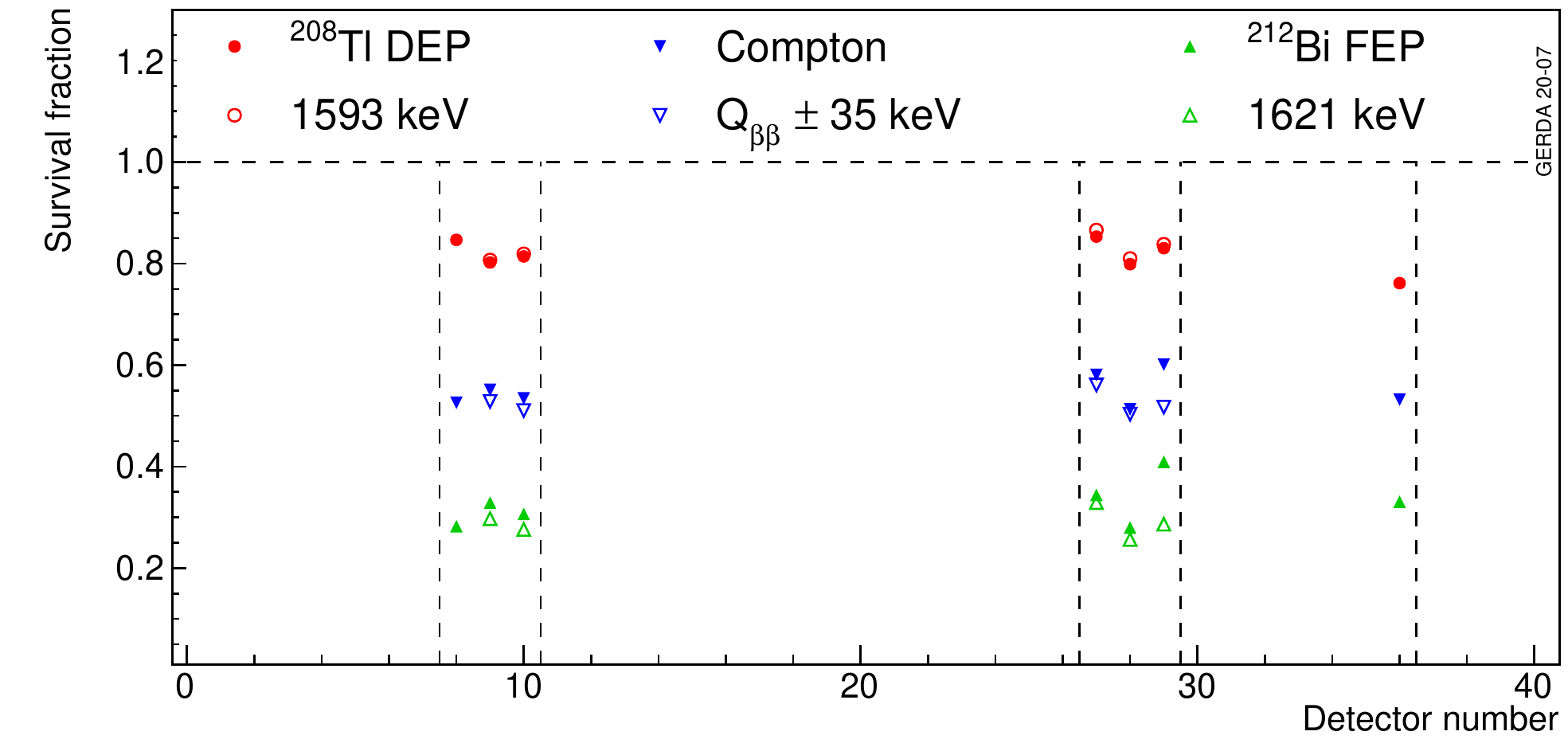}
    \caption{Survival fractions of events in the $^{208}$Tl~DEP, $^{212}$Bi~FEP and CC($Q_{\beta\beta}$) for each coaxial detector after ANN and risetime cuts. 
    Open (filled) symbols show the calibration dataset before (after) the 2018 upgrade. 
    The uncertainties are only statistical and smaller than the markers.}
    \label{fig:coax_sf}
\end{figure}

\begin{figure}[htb]
    \centering
    \includegraphics[width=0.48\textwidth]{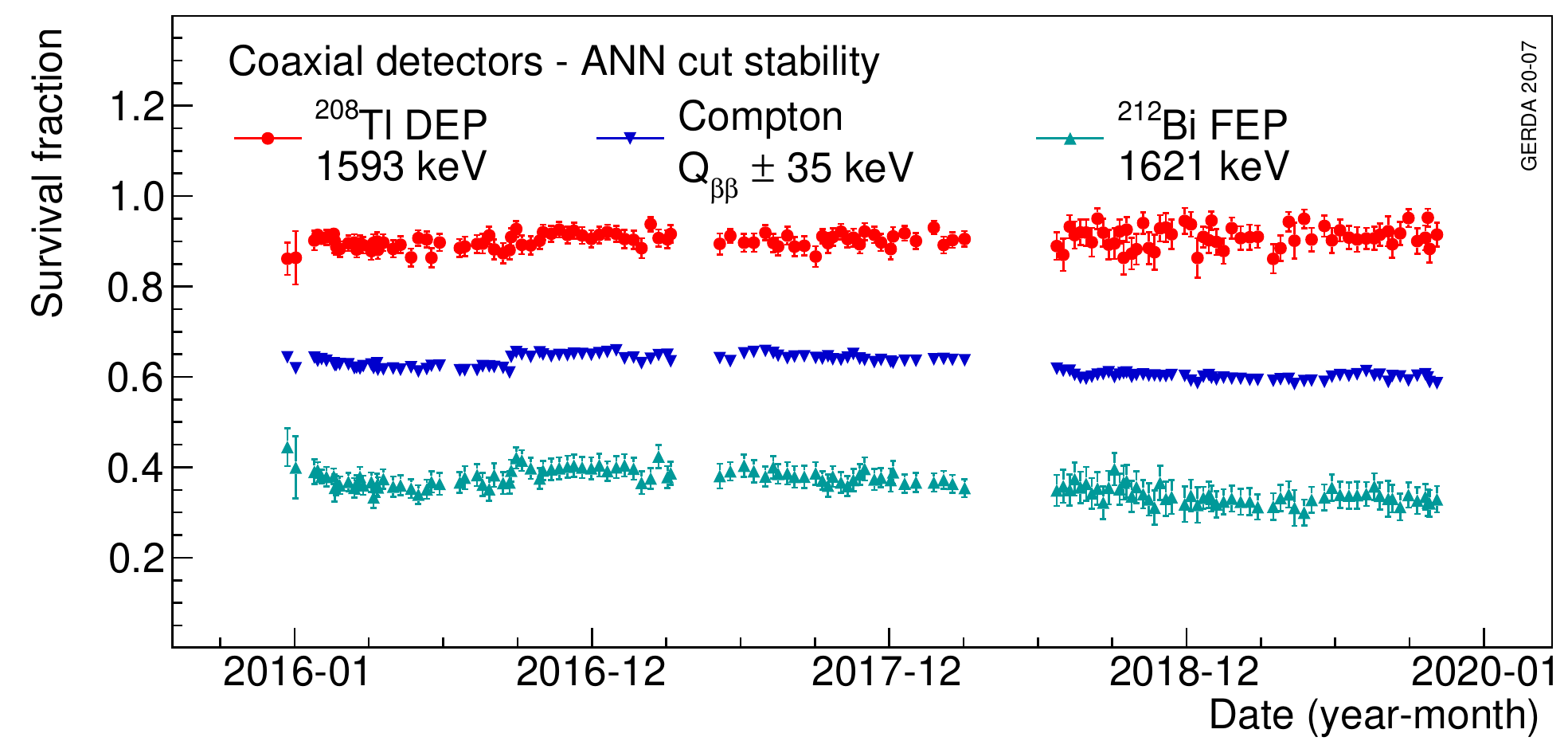}
    \caption{Average survival fractions of events after the ANN cut in the $^{208}$Tl~DEP, $^{212}$Bi~FEP and CC($Q_{\beta\beta}$) for coaxial detectors as a function of time. Each data point represents a calibration run with its statistical uncertainty.}
    \label{fig:ann_stability}
\end{figure}

\begin{figure}[htb]
    \centering
    \includegraphics[width=0.48\textwidth]{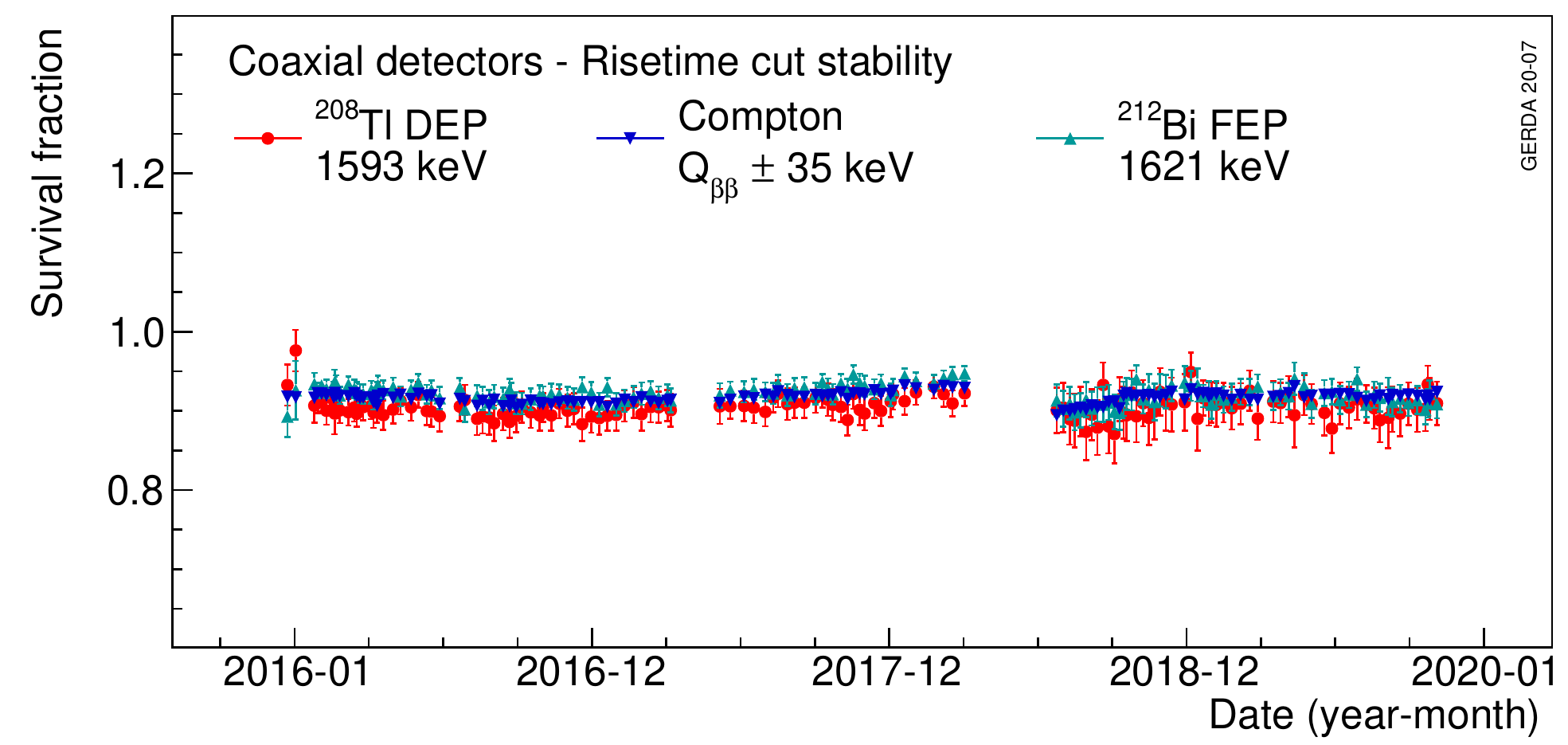}
    \caption{Average survival fractions of events after the risetime cut in the $^{208}$Tl~DEP, $^{212}$Bi~FEP and CC($Q_{\beta\beta}$) for coaxial detectors as a function of time. Each data point represents a calibration run with its statistical uncertainty.}
    \label{fig:rt_stability}
\end{figure}

Various survival fractions for all coaxial detectors, after applying both ANN and risetime cuts for the two data taking periods, are depicted in Fig.~\ref{fig:coax_sf}.
In addition, the stabilities of these cuts in three calibration energy regions are plotted in Figs.~\ref{fig:ann_stability} and \ref{fig:rt_stability}.
Apart from a slight improvement of the ANN rejection of the Compton continuum events at $Q_{\beta\beta}$ after the 2018 upgrade due to an improved signal cable management, an overall 3\% level stability in PSD performance is observed over the course of {\sc Gerda} Phase II.
Compared to {\sc Gerda} Phase I \cite{Agostini:2013jta}, on average a 7\% relative worsening is observed, mostly attributed to the different electronics scheme, hence different noise.
The risetime cut also shows a very stable behavior during this period.

\begin{figure}[htb]
    \centering
    \includegraphics[width=0.48\textwidth]{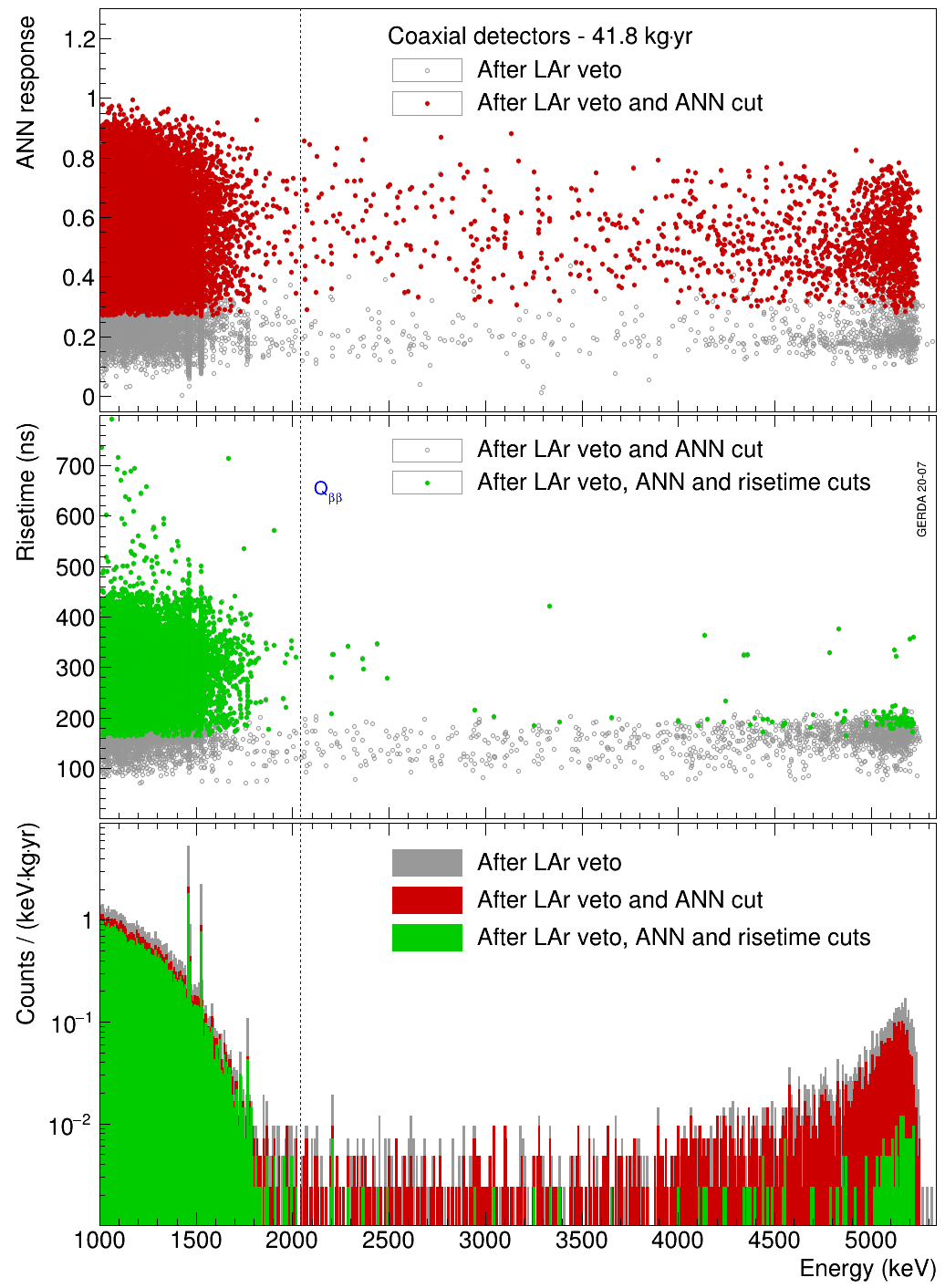}
    \caption{Top: ANN classifier of physics events from coaxial detectors as a function of energy. Middle: Risetime of events surviving the ANN cut as a function of energy. Bottom: Energy distribution of events from coaxial detectors before and after the ANN and risetime cuts.}
    \label{fig:coax_phys}
\end{figure}

The result of the pulse shape analysis applied on the full 41.8 kg$\cdot$yr physics data exposure, after the LAr veto cut, is summarized in Fig.~\ref{fig:coax_phys}.
The ANN is preserving 80.5(3)\% of the $2\nu\beta\beta$ decay event sample while removing 62(1)\% of the $^{40}$K and $^{42}$K lines.
The reduction of the $^{214}$Bi~FEPs at 1806~keV and 2204~keV is also visible.
However, it is in general inefficient at rejecting fast and degraded $\alpha$ events from about 3500~keV to 5500~keV, as 58.4(8)\% of them survive.
After applying also the risetime cut, 67.8(4)\% of $2\nu\beta\beta$ decay events remain.
The $\alpha$ background is suppressed by a factor of $\sim$23 above 3500~keV, only 93 events out of 2169 survive the PSD cuts.
The complementarity between the two cuts yields for the coaxial detectors a dataset that remains background-free in the region of interest with a background index of $5.0^{+2.6}_{-2.0}~\cdot~10^{-4}$~counts/(keV$\cdot$kg$\cdot$yr).

\section{Events with incomplete charge collection}
\label{sec:deltaE}

Events from the n$^+$ layer or the groove featuring slow or incomplete charge collection (see Fig. 2) have an uncertainty on energy reconstruction because the ZAC filter~\cite{Agostini:2015pta} is optimized for the FEP resolution with a relatively short integration time. 
These particular events, especially for coaxial detectors, can survive the ANN and risetime cuts.
In order to discard these events with uncertain ZAC energy reconstruction, an additional classification based on energies reconstructed with pseudo Gaussian filters with different integration times is performed.
The energy is reconstructed with an integration time of 4~$\mu$s ($E_{short}$) and 20~$\mu$s ($E_{long}$).
The ratio $E_{short}/E_{long}$ is then normalized to its average observed in events from the 2615~keV line in calibration data.
The classifier is defined as:
\begin{equation}
	\delta E = \left(  \frac{ E_{short}/E_{long}}{ \langle E_{short}/E_{long} \rangle_{FEP} } - 1  \right) \cdot E 
	\label{eq:dE}
\end{equation}
where $E$ is the default ZAC-reconstructed  energy for each event.
With this definition, the classifier $\delta E$ has an average of 0 keV.

The $E_{short}$/$E_{long}$ distribution normalization is performed time dependently in order to account for possible instabilities of the readout electronics: for each calibration run the mean of the fitted Gaussian to the FEP distribution is used as a normalization factor for the following physics events.

\begin{figure}[ht]
    \centering
    \includegraphics[width=0.45\textwidth]{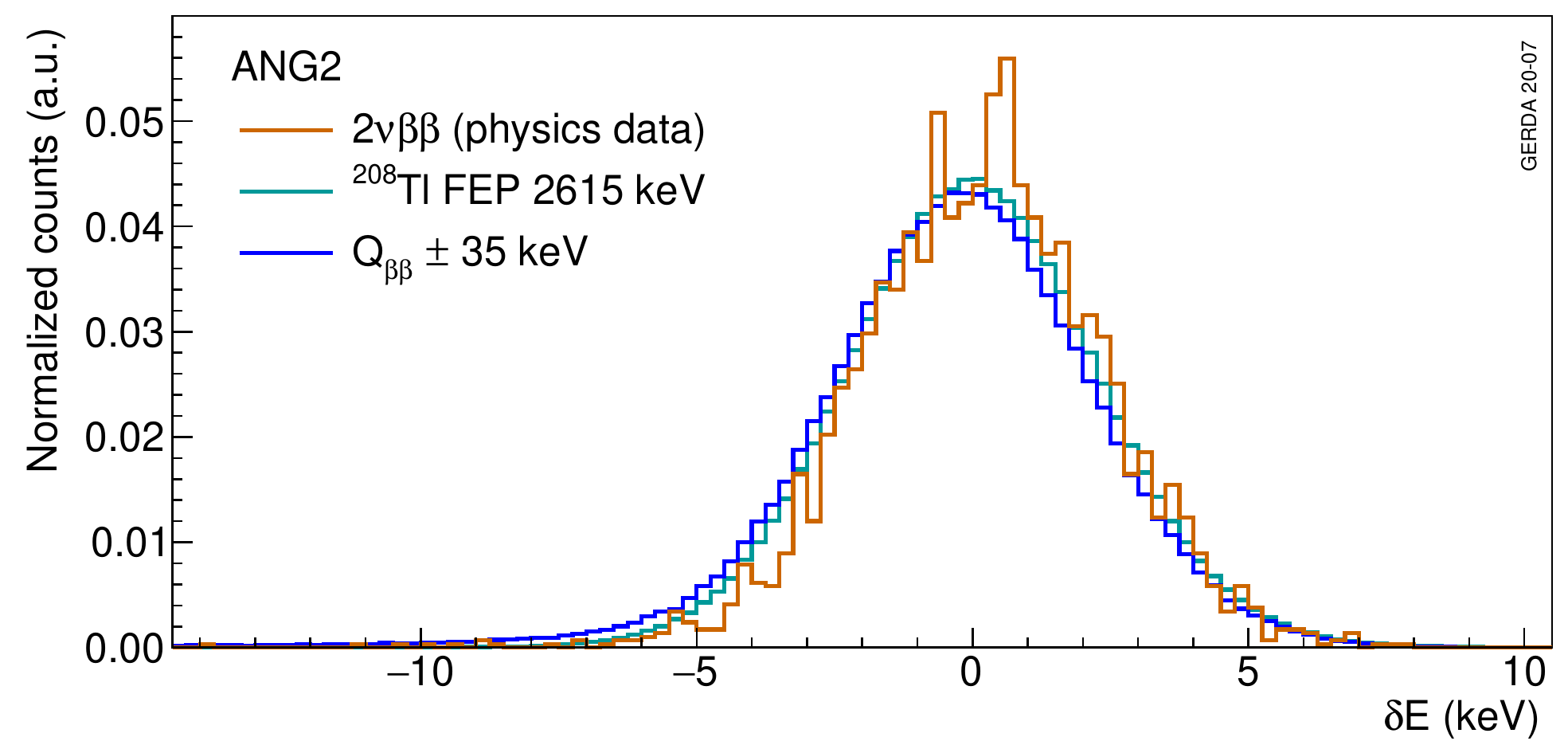}
    \caption{Distribution of the $\delta E$ classifier for calibration events in the $^{208}$Tl~FEP and in the Compton region around $Q_{\beta\beta}$ as well as for physics events in the $2\nu\beta\beta$ region for the coaxial ANG2 detector. The histograms are normalized to their integrals.}
    \label{fig:dE_distributions} 
\end{figure}

Figure~\ref{fig:dE_distributions} shows the resulting $\delta E$ distributions for calibration events in the $^{208}$Tl~FEP and in the CC($Q_{\beta\beta}$) as well as for physics events in the $2\nu\beta\beta$ region for the ANG2 detector as an example.
Events in the Compton continuum have a higher fraction with large negative $\delta E$ values.
This is due to the higher fraction of pulses with incomplete charge collection that is not present in a peak where the whole energy has been collected.
Physics data in the $2\nu\beta\beta$ region shows a similar Gaussian distribution as calibration data with no significant energy dependence.

A cut value is applied to the lower side of the $\delta E$ distribution and it is set for each detector separately $3\sigma$ away from zero.\footnote{No event with $\delta E$ classifier greater than $3\sigma$ was found in the background analysis window before $A/E$ or ANN+risetime cuts, therefore no high $\delta E$ cut was applied.}
The cut value is loose enough that more than 99\% of signal events are kept whereas those with uncertain energy (significant difference between energies reconstructed by the ZAC and Gauss algorithms) are rejected.

For the BEGe and IC detectors, most of the $2\nu\beta\beta$ decay events rejected by the $\delta E$ cut are also rejected by the $A/E$ cut.
For BEGe detectors out of 3477 events cut by either method only 7 are rejected by the $\delta E$ cut (for energies above 1000~keV 12 events in total are cut by $\delta E$).
For the IC detectors none of the 521 $2\nu\beta\beta$ decay events is removed by the $\delta E$ cut alone. 

For the coaxial detectors the correlation between ANN, risetime and $\delta E$ cuts is weaker.
Analyzing the $2\nu\beta\beta$ decay region for the full Phase II dataset, one gets 4970 events (out of 15433) removed by either method, 1 event cut by all three methods and 57 by $\delta E$ only. 
83 events from coaxial detectors with $E > 1000$~keV are cut by $\delta E$ only. Many of these events show slow pulses that could originate from the detector surface but are not cut by ANN or risetime cuts.

The survival fractions for different energy regions of physics and calibration data have been studied for each detector separately. 
Survival fractions of $2\nu\beta\beta$ decay events for the different detector types are presented in Table~\ref{tab:dE_cut} before and after the other PSD cuts.
Values for DEP, FEP, CC($Q_{\beta\beta}$) and $2\nu\beta\beta$ decay events for the two datasets before and after the 2018 upgrade are very close to 100\%. 
The Compton region as well as the $2\nu\beta\beta$ region show lower acceptance because of the contribution of slow pulses. 
 
For BEGe and IC detectors the impact of the $\delta E$ cut is negligible, while for coaxial detectors a small correction of the efficiency has to be taken into account.

\begin{table}[t]
\centering
\caption{\label{tab:dE_cut} Survival fractions of $2\nu\beta\beta$ decay events for the $\delta E$ cut without and in combination with other PSD methods ($A/E$ or ANN+risetime). The uncertainties are statistical only.}
\begin{tabular}{lcc}\cline{1-3} \noalign{\vskip\doublerulesep
         \vskip-\arrayrulewidth} \cline{1-3} 
               Detector & \multicolumn{2}{c}{Survival fraction [\%]}  \rule{0pt}{2.6ex}   \\
               type     & before PSD     & after PSD  \\ \hline
            Coaxial      & 99.57 $\pm$ 0.05 & 99.46  $\pm$ 0.07 \\
            BEGe         & 98.47 $\pm$ 0.09 & 99.96 $\pm$ 0.02 \\
            IC           & 98.58 $\pm$ 0.20 & $>99.90$ (95\% C.I.)  \\ 
            \cline{1-3} \noalign{\vskip\doublerulesep
         \vskip-\arrayrulewidth} \cline{1-3} 
\end{tabular}
\end{table}


\section{PSD detection efficiencies at Q$_{\beta\beta}$}
\label{sec:efficiency}

\begin{figure}
\centering
\includegraphics[width=0.45\textwidth]{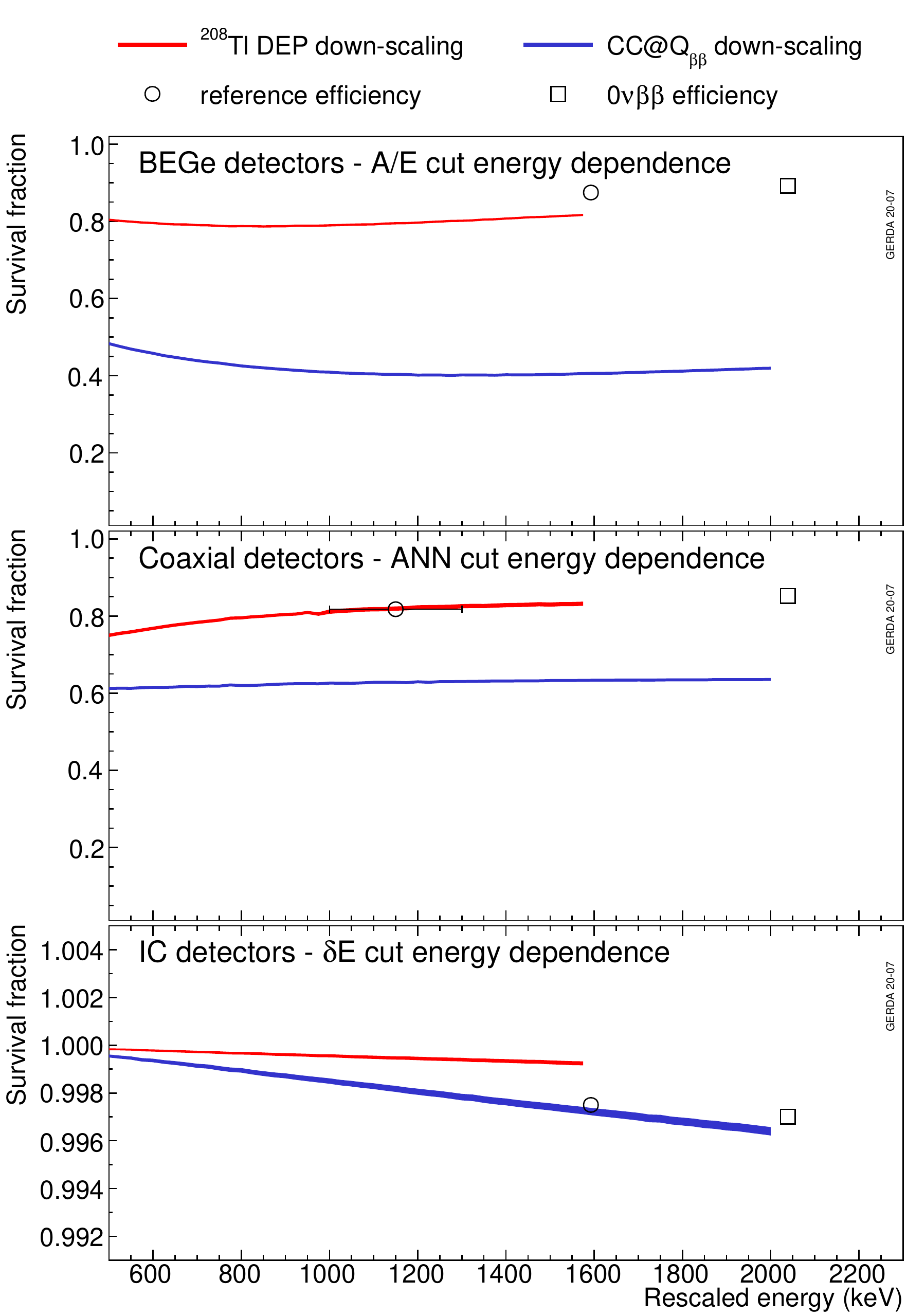}
\caption{Extrapolation of reference survival fractions (circles) to $Q_{\beta\beta}$ (squares) using the energy dependence deduced from indicated two samples of down-scaled waveforms (see text). Examples are given for BEGe (top), coaxial (middle) and IC detectors (bottom) for the $A/E$, ANN and $\delta E$ methods, respectively. 
As to the BEGe\,-\,$A/E$ example on top: the reference survival fraction is corrected for Compton events below the $^{208}$Tl~DEP while for the rescaled distribution (red) this correction is missing.}
\label{fig:rsc-samples} 
\end{figure}

In the absence of signal proxies at $Q_{\beta\beta}$ in calibration and physics data and due to a not sufficiently accurate modeling of the pulse shape response in simulation,  
$\varepsilon_{\rm{PSD}}$ is estimated for point contact and coaxial detectors from the extrapolation of the survival fractions of the $^{208}$Tl~DEP and of the $2\nu\beta\beta$ decay events at 1593 keV and 1150 keV to $Q_{\beta\beta}$ (see circles and squares in Fig. \ref{fig:rsc-samples}).

\begin{table*}
	\centering
    \caption{Average signal detection efficiencies at $Q_{\beta\beta}$ of individual PSD methods for the different detector types and data taking periods.
    $\varepsilon_{\tiny{\textrm{PSD}}}$ is estimated via a MC sampling of individual values (see Table \ref{tab:det-eff}) and is thus different from the product of the individual PSD methods efficiencies reported here.}
	\label{tab:final_eff}
	\begin{tabular}{lcccccc}
	    \cline{1-7} \noalign{\vskip\doublerulesep
         \vskip-\arrayrulewidth} \cline{1-7} 
	    & \multicolumn{2}{c}{Dec 2015-May 2018} & & \multicolumn{3}{c}{July 2018-Nov 2019} \rule{0pt}{2.6ex} \\
	    \cline{2-3} \cline{5-7}
        & Coaxial & BEGe && Coaxial & BEGe & Inverted coaxial \rule{0pt}{2.6ex} \\  \hline 
        Exposure $\mathcal{E}$ & 28.6~kg$\cdot$yr & 31.5~kg$\cdot$yr && 13.2~kg$\cdot$yr & 21.9~kg$\cdot$yr & 8.5~kg$\cdot$yr \\
		$\varepsilon_{\tiny{\textrm{ANN}}}$ &  82.5\% & - && 81.8\% & - & - \\ 
		$\varepsilon_{\tiny{\textrm{risetime}}}$ & 85.7\% & - && 85.0\% & - & -  \\
		$\varepsilon_{\tiny{\textrm{A/E}}}$ & - & 88.4\% && - & 89.3\% & 90.0\% \\
		$\varepsilon_{\tiny{\textrm{$\delta E$}}}$ & 99.5\% & 100.0\% && 99.7\% & 100.0\% & 99.7\% \\
		$\varepsilon_{\tiny{\textrm{PSD}}}$ & \textbf{(69.1 $\pm$ 5.6)\%} & \textbf{(88.2 $\pm$ 3.4)\%} && \textbf{(68.8 $\pm$ 4.1)\%} & \textbf{(89.0 $\pm$ 4.1)\%} & \textbf{(90.0 $\pm$ 1.8)\%} \\ 
	    \cline{1-7} \noalign{\vskip\doublerulesep
         \vskip-\arrayrulewidth} \cline{1-7} \\
	\end{tabular}
\end{table*}

For that purpose, sets of down-scaled waveforms are used to evaluate the energy dependence of the various PSD methods.
This is based on the observation that the energy dependence of PSD classifiers is dominated by the electronic noise in {\sc Gerda}.
Two samples from calibration data have been considered, namely $^{208}$Tl~DEP and CC($Q_{\beta\beta}$) events (see Fig. \ref{fig:calspectrum}).
The waveform $\omega'(k,E)$ at energy $E$ is produced by down-scaling the waveform $\omega(k,E_S)$, of true energy $E_S$, from one of these two samples and superimposing it to a baseline sample $b(k)$ obtained from random triggers:
\begin{equation}
	\omega'(k,E) = \omega(k,E_S) \cdot \frac{E}{E_S} + b(k) \cdot \left(\sqrt{1-\left(\frac{E}{E_S}\right)^2}\right)
	\label{eq:rsc-wf}
\end{equation}
where $k$ is the index running from 0 $\mu$s to 160 $\mu$s and from 0 $\mu$s to 10 $\mu$s for the low and high frequency traces introduced in Sec. \ref{sec:pulses}, respectively.
By applying the complete set of digital processing steps and analysis methods described earlier, such a procedure allows to estimate the contribution of the energy dependence, induced by the signal-to-noise ratio variation, on the PSD methods results (for more details see \cite{Lazzaro:2019pda}).

Figure~\ref{fig:rsc-samples} shows the survival fractions of the $^{208}$Tl~DEP and CC($Q_{\beta\beta}$) datasets as a function of the down-scaled energy for the $A/E$, ANN and $\delta E$ methods\footnote{The risetime method is not shown as its survival fraction is insensitive to the signal-to-noise ratio at these energies.}.
Each plot depicts the energy dependence of a given detector type, as a result of a weighted average of the individual detector responses, to illustrate the general trends.  
The reference efficiency is extrapolated to $Q_{\beta\beta}$ using the average energy dependence of $^{208}$Tl~DEP and CC($Q_{\beta\beta}$) datasets\footnote{Energy dependencies are calculated from [1050,\,1550]~keV and [1550,\,1900]~keV for the $^{208}$Tl~DEP and CC($Q_{\beta\beta}$) datasets, respectively.}.
Table \ref{tab:final_eff} shows how the different PSD procedures contribute to the resulting average detection efficiencies $\varepsilon_{\rm{PSD}}$ of the various detector types for each data taking period. 
The efficiencies of the point contact detectors are almost 30\% larger than those of the coaxial detectors. 

The analysis of the final {\sc Gerda} results \cite{Agostini:2020prl} is based on time-dependent and detector-wise datasets.
Table \ref{tab:det-eff} (Appendix A) reports the overall PSD efficiency for each detector separately. 
Since the central values of the individual detector efficiencies exhibit significant shifts, the average overall PSD efficiencies $\varepsilon_{\rm{PSD}}$ of Table \ref{tab:final_eff} have been obtained, in fact, from an exposure weighted Monte-Carlo sampling, and not from a simple average.

The systematic uncertainty of the extrapolation is estimated from the difference of the slopes of the two down-scaled waveform samples.
It is on average about 0.5\% and 1.3\% for point contact and coaxial detectors, respectively,
the latter detector type being more sensitive to the noise due to its larger p$^{+}$ contact capacitance.

In addition, two other systematic effects have been taken into account: 1) the difference between the calibration and physics data and 2) the difference between signal proxies and $0\nu\beta\beta$ decay events.
The former applies only to point contact detectors as the $^{208}$Tl~DEP from calibration data is used as signal proxy.
Shifting the $A/E$ cut, for each detector, by the $A/E$ distribution bias observed between the $^{208}$Tl~DEP and $2\nu\beta\beta$ decay (Fig. \ref{fig:aoe_compare}) leads to an average relative systematic uncertainty of 1.9\%.
The latter makes use of pulse shape simulation (see \ref{app:pss}) to quantify the PSD performance bias between $2\nu\beta\beta$ decay and $^{208}$Tl~DEP events and those coming from $0\nu\beta\beta$ decay.
Indeed, the signal proxies feature much lower energies hence different Bremsstrahlung contribution.
In addition, $^{208}$Tl~DEP events have a higher probability to happen close to the detector surface while $0\nu\beta\beta$ decays would occur homogeneously throughout the detector bulk.
A 2.3\% and 1.5\% absolute uncertainty has been estimated for BEGe and IC detectors, respectively, while it amounts to 4\% for coaxial detectors.
This last estimate is larger due to the difficulty to match the ANN training performance in simulation with the data.

\section{Summary}
\label{sec:summary}

Nowadays, running a background-free $0\nu\beta\beta$ decay experiment is essential to boost the $T^{0\nu}_{1/2}$ sensitivity on a reasonable time scale.
Over the past years, the {\sc Gerda} collaboration demonstrated the feasibility of such a program by upgrading its initial setup with additional point contact detectors (BEGe and IC), a LAr veto instrumentation and lower mass holders.
As a consequence, the sensitivity linearly increased with the exposure.
The interplay between passive and active shielding techniques has proven to be highly effective.
In this paper, we focused on the ability offered by germanium detectors to analyze the topological structure of the recorded events.
This topology, distinct for signal-like and background-like events at a 100 ns time scale, is best scrutinized with the high frequency based data acquisition system of {\sc Gerda}.
Using the Phase II dataset, we confirmed the superior discriminating power of point contact detectors (BEGe and IC) over the historical coaxial detectors thanks to the simple $A/E$ parameter.
Their $0\nu\beta\beta$ decay PSD efficiency is 89\% and 69\%, respectively for a similar background index of about $5\cdot10^{-4}$~counts/(keV$\cdot$kg$\cdot$yr) after all cuts.
We also demonstrated the high and stable performance in LAr of the newly produced enriched inverted-coaxial point contact germanium detectors.
The \textsc{LEGEND} collaboration will deploy these IC detectors for its future $^{76}$Ge $0\nu\beta\beta$ decay search program.

The total detection efficiencies for $0\nu\beta\beta$ decay are discussed and listed for each detector 
in Appendix A.

\begin{acknowledgements}
The {\sc Gerda} experiment is supported financially by
the German Federal Ministry for Education and Research (BMBF),
the German Research Foundation (DFG),
the Italian Istituto Nazionale di Fisica Nucleare (INFN),
the Max Planck Society (MPG),
the Polish National Science Centre (NCN),
the Foundation for Polish Science (TEAM/2016-2/17),
the Russian Foundation for Basic Research,
and the Swiss National Science Foundation (SNF).
This project has received funding/support from the European Union's
\textsc{Horizon 2020} research and innovation programme under
the Marie Sklodowska-Curie grant agreements No 690575 and No 674896.
This work was supported by the Science and Technology Facilities Council, 
part of the U.K. Research and Innovation (Grant No. ST/T004169/1). 
J. Huang and C. Ransom thank the UZH for the Postdoc and Candoc
Forschungskredit fellowships respectively.
The institutions acknowledge also internal financial support.
The {\sc Gerda} collaboration thanks the directors and the staff of LNGS for 
their continuous strong support of the {\sc Gerda} experiment.
\end{acknowledgements}

\appendix

\section{Efficiency tables per detector}
\label{app:tables}

The final results of {\sc Gerda} on the search for $0\nu\beta\beta$ decay \cite{Agostini:2020prl} have been obtained from data partitions, defined detector-wise by stable data taking periods, that notably include the stability of PSD performance. Table \ref{tab:det-eff} lists for each germanium detector, before and after the 2018 upgrade, the PSD cut efficiencies $\varepsilon_{\rm{PSD}}$ as well as  the total efficiencies $\varepsilon_{0\nu\beta\beta}$  of detecting $0\nu\beta\beta$ decays the latter entering Eq. (1) of \cite{Agostini:2020prl}:
\begin{equation}
	\varepsilon_{0\nu\beta\beta} = f_{76} \cdot f_{\rm{av}} \cdot \varepsilon_{\rm{fep}} \cdot \varepsilon_{\rm{PSD}} \cdot \varepsilon_{\rm{LAr}} .
	\label{eq:tot-effi}
\end{equation}
For the coaxial and BEGe detectors, Table S1 of the Supplementary Material of \cite{Agostini:2019hzm} provides the $^{76}$Ge enrichment fractions $f_{76}$ and the active volume fractions  $f_{\rm{av}}$.
The electron containment efficiencies $\varepsilon_{\rm{fep}}$ of the period before the upgrade are also provided there while a new computation has been used for the period after the upgrade that is reported in Table \ref{tab:det-eff}.
Table \ref{tab:det-av} shows the active volume fractions for the IC detectors. 
The efficiencies $\varepsilon_{\rm{LAr}}$ of the LAr veto cut for the two data taking periods are 0.977(1) and 0.982(1), respectively \cite{Agostini:2020prl}. Equation \ref{eq:tot-effi} deliberately neglects the efficiencies of muon veto and quality cuts both being larger than 0.999.
The uncertainty of all efficiencies are incorporated via a Monte-Carlo sampling from which we retrieve the $\varepsilon_{0\nu\beta\beta}$ central value and its standard deviation.

\begin{table*}
\centering
    \caption{Detector-wise PSD cut efficiencies $\varepsilon_{\tiny{\textrm{PSD}}}$ and  the total detection efficiencies $\varepsilon_{0\nu\beta\beta}$  for $0\nu\beta\beta$ decays used in the final analysis of \cite{Agostini:2020prl} and corresponding exposures $\mathcal{E}$. For the period after the 2018 upgrade also the electron containment efficiencies $\varepsilon_{\rm{fep}}$ are listed (see text). Quoted uncertainties account for statistics and systematics.
    The channels with empty entries were used in anti-coincidence only.}
    \label{tab:det-eff}
    \begin{tabular}{ccccccccccc}
        \cline{1-10} 
        \noalign{\vskip\doublerulesep\vskip-\arrayrulewidth}
        \cline{1-10} 
        \multicolumn{5}{c}{Dec 2015-May 2018} & &		\multicolumn{4}{c}{July 2018-Nov 2019} \rule{0pt}{2.6ex} \\ \cline{1-5}  \cline{7-10}
		\multirow{2}{*}{\#} & Detector & $\mathcal{E}$ & \multirow{2}{*}{$\varepsilon_{\tiny{\textrm{PSD}}}$} & \multirow{2}{*}{$\varepsilon_{0\nu\beta\beta}$} & & \multirow{2}{*}{$\varepsilon_{\tiny{\textrm{fep}}}$} & $\mathcal{E}$ & \multirow{2}{*}{$\varepsilon_{\tiny{\textrm{PSD}}}$} & \multirow{2}{*}{$\varepsilon_{0\nu\beta\beta}$} \rule{0pt}{2.6ex} \\
		& label & kg.yr & & & & & kg.yr & & \\ \hline
0  & GD91A & 1.167 & 0.890(29) & 0.610(25) & &0.895(2) & 0.736 & 0.891(23) & 0.612(22) \\
1  & GD35B & 1.455 & 0.895(27) & 0.634(24) & &0.899(2) & 0.880 & 0.891(30) & 0.632(26) \\
2  & GD02B & 1.181 & 0.878(22) & 0.598(21) & &0.891(2) & 0.733 & 0.885(22) & 0.603(21) \\
3  & GD00B & 1.193 & 0.859(27) & 0.583(24) & &0.893(2) & 0.753 & 0.885(21) & 0.601(21) \\
4  & GD61A & 1.340 & 0.876(21) & 0.606(21) & &0.899(2) & 0.858 & 0.901(22) & 0.624(21) \\
5  & GD89B & 0.639 & 0.869(21) & 0.572(21) & &0.885(2) & 0.707 & 0.894(22) & 0.587(22) \\
6  & GD02D & -     & -         & -         & & -       & -     & -         & -         \\
7  & GD91C & 0.234 & 0.867(23) & 0.592(22) & &0.892(2) & 0.736 & 0.894(29) & 0.611(25) \\
8  & ANG5  & 5.067 & 0.760(32) & 0.486(37) & & -       & -     & -         & -         \\
9  & RG1   & 3.914 & 0.670(31) & 0.465(39) & &0.919(2) & 2.392 & 0.661(32) & 0.463(39) \\
10 & ANG3  & 4.557 & 0.653(29) & 0.449(39) & &0.920(2) & 2.711 & 0.681(32) & 0.472(40) \\
11 & GD02A & 1.016 & 0.904(23) & 0.621(21) & &0.889(2) & 0.640 & 0.886(24) & 0.609(22) \\
12 & GD32B & 1.257 & 0.909(95) & 0.620(67) & &0.890(2) & 0.780 & 0.90(20) & 0.62(13)\\
13 & GD32A & 0.489 & 0.865(22) & 0.583(23) & &0.883(2) & 0.529 & 0.886(34) & 0.597(29) \\
14 & GD32C & 1.377 & 0.878(23) & 0.609(21) & &0.898(2) & 0.872 & 0.898(57) & 0.624(42) \\
15 & GD89C & 1.056 & 0.870(22) & 0.581(23) & &0.880(2) & 0.698 & 0.894(27) & 0.597(25) \\
16 & GD61C & 1.072 & 0.874(28) & 0.594(24) & &0.889(2) & 0.734 & 0.883(24) & 0.602(22) \\
17 & GD76B & 0.698 & 0.835(21) & 0.538(20) & &0.879(2) & 0.426 & 0.858(22) & 0.553(21) \\
18 & GD00C & 1.479 & 0.892(21) & 0.618(22) & &0.890(2) & 0.956 & 0.902(22) & 0.618(22) \\
19 & GD35C & 1.202 & 0.916(39) & 0.635(30) & &0.889(2) & 0.739 & 0.895(44) & 0.620(34) \\
20 & GD76C & 1.371 & 0.902(43) & 0.614(32) & &0.899(2) & 0.964 & 0.911(47) & 0.621(35) \\
21 & GD89D & 0.945 & 0.879(32) & 0.576(26) & &0.880(2) & 0.617 & 0.891(51) & 0.584(37) \\
22 & GD00D & 1.550 & 0.889(21) & 0.613(21) & &0.898(2) & 0.954 & 0.902(31) & 0.622(26) \\
23 & GD79C & 1.206 & 0.877(21) & 0.596(20) & &0.896(2) & 0.953 & 0.905(21) & 0.615(20) \\
24 & GD35A & 1.460 & 0.882(20) & 0.618(21) & &0.901(2) & 0.901 & 0.896(22) & 0.630(22) \\
25 & GD91B & 0.420 & 0.876(50) & 0.601(37) & &0.893(2) & 0.734 & 0.879(35) & 0.603(28) \\
26 & GD61B & 1.172 & 0.896(22) & 0.615(21) & &0.895(2) & 0.808 & 0.899(21) & 0.616(21) \\
27 & ANG2  & 4.750 & 0.743(32) & 0.504(41) & &0.924(2) & 2.751 & 0.740(31) & 0.508(40) \\
28 & RG2   & 4.017 & 0.630(30) & 0.400(34) & &0.918(2) & 2.542 & 0.640(27) & 0.412(32) \\
29 & ANG4  & 4.521 & 0.683(32) & 0.477(40) & &0.920(2) & 2.783 & 0.707(31) & 0.498(39) \\
30 & GD00A & 0.945 & 0.885(22) & 0.598(22) & &0.884(2) & 0.582 & 0.864(24) & 0.584(22) \\
31 & GD02C & 1.462 & 0.899(22) & 0.618(21) & &0.898(2) & 0.925 & 0.904(21) & 0.623(21) \\
32 & GD79B & 0.861 & 0.896(39) & 0.608(31) & &0.894(2) & 0.469 & 0.900(24) & 0.612(23) \\
33 & GD91D & 1.003 & 0.858(44) & 0.588(34) & &0.895(2) & 0.813 & 0.894(44) & 0.614(34) \\
34 & GD32D & 1.232 & 0.874(23) & 0.617(22) & &0.897(2) & 0.775 & 0.873(25) & 0.618(23) \\
35 & GD89A & 0.976 & 0.864(21) & 0.585(21) & &0.888(2) & 0.591 & 0.883(52) & 0.597(39) \\
36 & ANG1  & 1.812 & 0.649(33) & 0.403(36) & & -       & -     & -         & -         \\
37 & IC50B & -     & -         & -         & &0.920(2) & 2.213 & 0.894(18) & 0.664(14) \\
38 & IC48A & -     & -         & -         & &0.921(2) & 2.159 & 0.910(18) & 0.676(14) \\
39 & IC50A & -     & -         & -         & &0.910(2) & 1.814 & 0.896(17) & 0.646(14) \\
40 & IC74A &       &           &           & &0.921(2) & 2.398 & 0.900(15) & 0.652(12) \\
    \cline{1-10} 
    \noalign{\vskip\doublerulesep\vskip-\arrayrulewidth} 
    \cline{1-10} 
    \end{tabular}
\end{table*}

\begin{table*}
\centering
    \caption{Active volume fractions $f_\textrm{av}$ of IC detectors used in the final analysis of \cite{Agostini:2020prl}. The $f_\textrm{av}$ values are pulled from a preliminary analysis found in \cite{Miloradovic:2019phd}. Quoted uncertainties account for statistics and systematics.}
    \label{tab:det-av}
    \begin{tabular}{cccccccc}
        \cline{1-3} 
        \noalign{\vskip\doublerulesep\vskip-\arrayrulewidth}
        \cline{1-3} 
        \multicolumn{3}{c}{IC detectors} \rule{0pt}{2.6ex} \\ \cline{1-3}
		\multirow{2}{*}{\#} & Detector & \multirow{2}{*}{$f_\textrm{av}$} \rule{0pt}{2.6ex} \\
		 & label & \\ \hline
        36 & IC48B & 0.935(5) \\
        37 & IC50B & 0.938(4) \\
        38 & IC48A & 0.936(7) \\
        39 & IC50A & 0.920(4) \\
        40 & IC74A & 0.913(6) \\
    \cline{1-3} 
    \noalign{\vskip\doublerulesep\vskip-\arrayrulewidth} 
    \cline{1-3} 
    \end{tabular}
\end{table*}

\section{Pulse shape simulation}
\label{app:pss}

\begin{figure}[ht]
\centering
\includegraphics[width=0.45\textwidth]{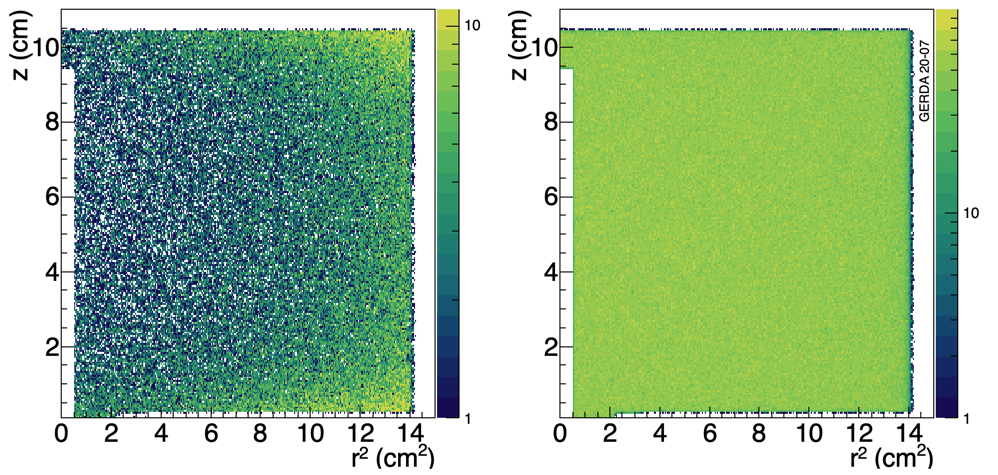}
\caption{Events distributions across the detector bulk from simulated $^{208}$Tl~DEP (left) and $0\nu\beta\beta$ decay (right) events.
The position of an event corresponds to the barycentre of individual energy depositions.}
\label{fig:pss-hit-pos} 
\end{figure}

The analysis of {\sc Gerda} is data-driven for all PSD methods thanks to the selection of relevant energy regions for signal proxies.
However, this comes at the expense of neglecting the specific decay dynamics at $Q_{\beta\beta}$.
For instance, the amount of Bremsstrahlung at 2039 keV is larger than at 1593~keV, hence a potential $0\nu\beta\beta$ decay event has a different $A/E$ ratio compared to $^{208}$Tl~DEP.
Also, $^{208}$Tl~DEP events do happen on average closer to the detector surface while $0\nu\beta\beta$ decay are homogeneously distributed all over the detector bulk (see Fig. \ref{fig:pss-hit-pos}).
This results in a larger event fraction of the latter population above the p$^{+}$ contact with higher $A/E$ values (see Fig. \ref{fig:wpot}).

These effects are best studied with a Monte-Carlo simulation of the {\sc Gerda} experiment and subsequent pulse shape simulation (PSS) of individual events occurring in the germanium detectors.
For this purpose, the \textsc{Geant4} based simulation package (\textsc{MaGe} \cite{Boswell:2010mr}) and detector configuration used for the background model of {\sc Gerda} \cite{Abramov:2019hhh} was employed to generate $^{208}$Tl and $^{212}$Bi decays in order to reproduce the calibration energy spectrum of each germanium detector.
In addition, $2\nu\beta\beta$ decays were simulated homogeneously in each detector, to cross-check the PSD efficiencies in simulation with the data, as well as $0\nu\beta\beta$ decays.
The energy spectrum of the two electrons emitted in the double-$\beta$ decay was sampled according to the distribution given in reference \cite{TRETYAK199543} implemented in Decay0 \cite{MC:double-beta}.

The individual hit positions and deposited energies are used to calculate the corresponding induced charge flow in the detector by means of electrostatic simulation software (ADL \cite{Bruyneel:2016zih} and Fieldgen \cite{Radford:2015}).
Each generated pulse is then convoluted by an optimized electronics response model (bi-quad filter \cite{Panas:2018phd}) of the {\sc Gerda} setup before taking into account a realistic gain and offset.
Subsequently, waveforms from baseline events, recorded during the physics data acquisition to estimate the random coincidence probability, are randomly picked-up and added to the convoluted pulses to emulate the physical signal-to-noise ratio.
Finally, the post-processing is identical to the PSD methods described in Sec. \ref{sec:bege} and \ref{sec:coax} to retrieve the $A/E$, ANN and risetime classifiers.

\begin{figure}[ht]
\centering
\includegraphics[width=0.45\textwidth]{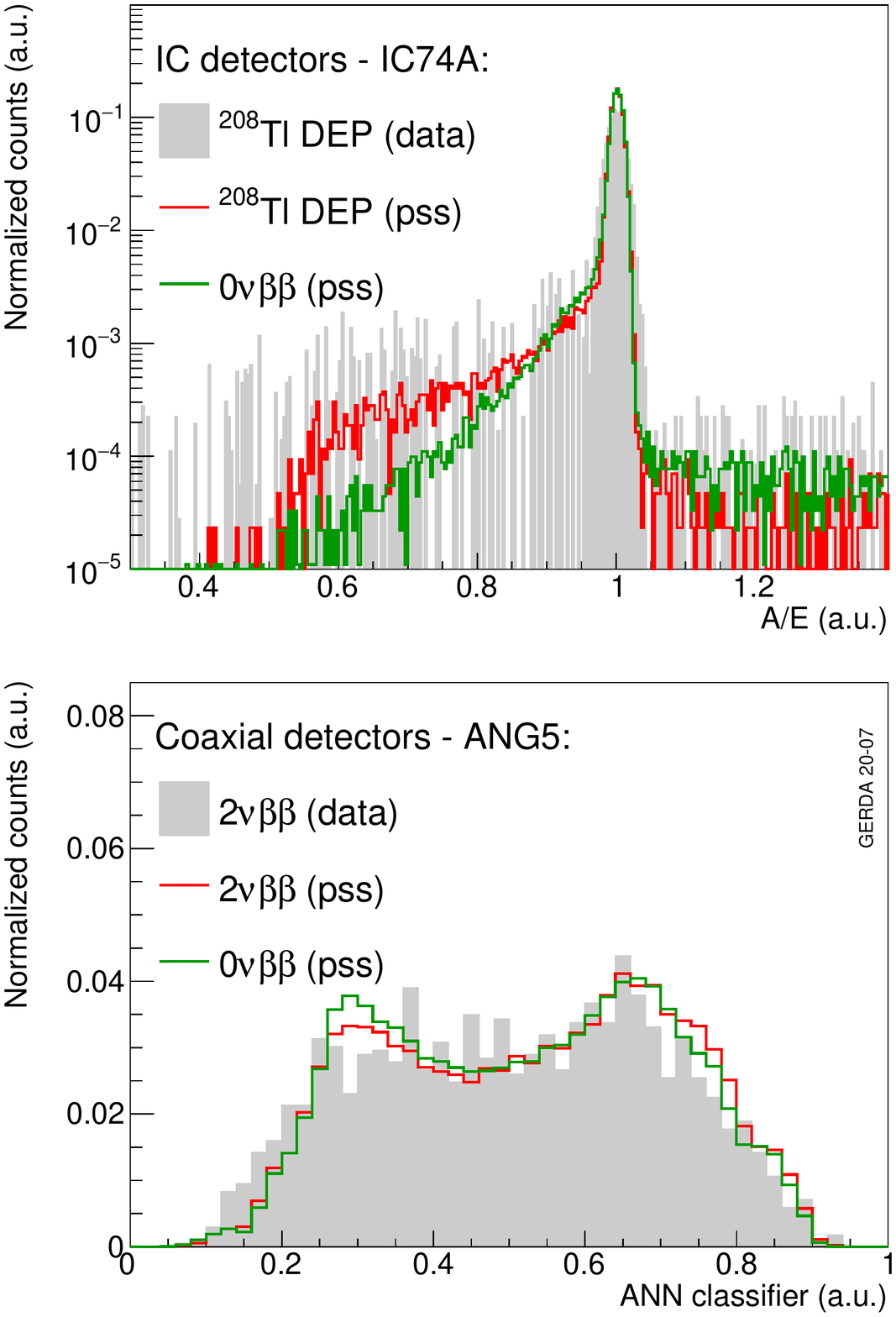}
\caption{Signal proxy and $0\nu\beta\beta$ decay events PSD classifier distributions for an inverted coaxial (top) and a coaxial (bottom) detector. 
The pulse shape simulation results (colored lines) are superimposed on the data (grey area).
The histograms are normalized to their integrals.}
\label{fig:pss-classifier} 
\end{figure}

In Fig. \ref{fig:pss-classifier}, the PSD classifier distribution of the simulation is compared to the data.
For the particular case of coaxial detectors, two approaches have been studied for the ANN analysis.
First, the training and cut values from the data were applied to the simulation.
Second, the simulated pulses were fed into the ANN for its training and the cuts were computed such that 90\% of simulated $^{208}$Tl~DEP events survive.
Due to the not accurate enough modeling of the {\sc Gerda} electronics response, the latter method was found to give results in better agreement with the data.

For IC detectors, the final systematic uncertainty from these data and simulation comparison is estimated by computing the survival fraction difference between $^{208}$Tl~DEP and $0\nu\beta\beta$ decay for each individual detector and then average it per detector type.
For coaxial detectors, the agreement of the $2\nu\beta\beta$ decay survival fractions between data and simulation is used to quantify the systematic uncertainty on the $0\nu\beta\beta$ signal efficiency.
The corresponding  BEGe detectors systematic uncertainty was estimated in previous studies \cite{Wagner:2017phd}.

\section{ANN for $\alpha$ event rejection}
\label{app:alpha}

In order to mitigate the high energy $\alpha$ background, an analysis based on a second ANN was developed at the beginning of {\sc Gerda} Phase II \cite{Agostini:2017iyd,Agostini:2018tnm}.
Other than in the ANN analysis described in Sec. \ref{sec:coax}, the neural network applied to $\alpha$ events was trained for different datasets using the highest possible statistics available~\cite{Kirsch:2014phd}.
For the training, the $2\nu\beta\beta$ decay events were used as signal-like sample, while $\alpha$-induced events did serve as a background-like sample. 
From the training point of view, all detectors had poor statistics for high energy events. 
Therefore, the training of the ANN-$\alpha$ has been performed with the whole available background dataset to optimize the selection. 
Later, two changes have been introduced into the $\alpha$ cut: i) in order to avoid any conflict with the determination of the survival fraction for the $2\nu\beta\beta$ decay events, calibration events at $Q_{\beta\beta} \pm 10$~keV, were provided as a signal-like sample instead, ii) input variables between 10\% and 90\% of the original trace were used for the supervised learning, what helped to reduce energy dependence of the ANN-$\alpha$ cut. 

Even though the statistics in the training data sets was much smaller compared to the MSE-based training, a very efficient -- and even better -- separation from the signal event class has been achieved. 
The discriminating parameter was set such that for physics data with $E > 3500$~keV a survival fraction of 10\% yielded, while retaining, on average, $\sim$95\% of $^{208}$Tl~DEP events in the calibration data. 
Only two detectors, RG1 and ANG4, featured slightly smaller $^{208}$Tl~DEP survival fractions of about 92\%. 
The corresponding statistical uncertainty on the 10\% cut on $\alpha$ events was in the range of (2 -- 3)\%. 

For reliability and technical reasons, the original TMVA algorithm used in~\cite{Kirsch:2014phd,Agostini:2017iyd,Agostini:2018tnm} called \textit{TMlpANN} (own ROOT ANN TMultiLayerPercetron class) was replaced by the recommended and supported \textit{MLPBFGS} (ANN Multilayer Perceptron class) in 2018, both using the same Broyden-Fletcher-Goldfarb-Shanno algorithm.
Two existing caveats persisted in the analysis.
First, the low statistics of the background sample ($\sim$50--100 events), was not sufficient to test the ANN from the beginning.
The new algorithm required to split the data in training and testing samples hence an even lower background statistics, which led to non-converging training in some detectors.
Second, few events with very fast signals, i.e. short risetime, were found to survive the ANN-$\alpha$ cut in the region of interest while they were thought to be $\alpha$ events with high confidence level.
This put in the question of reliability and robustness of this analysis approach, hence the decision to switch to the simpler risetime cut at the cost of a $\sim$10\% loss in signal efficiency.

\section{Projective likehood analysis}
\label{app:pl}

An alternative analysis strategy was pursued for the coaxial detectors using a projective likelihood (PL) to remove both MSEs and $\alpha$ background at once.
In this analysis the high-frequency waveform from the preamplifier output was smoothed by averaging every 5 adjacent samples, differentiated and then the maximum current value was found along with its corresponding time ($t_0$). 
For the analysis, 15 amplitudes before and 15 after $t_0$ were extracted from the original trace (31 in total). 
In order to further reduce the number of input variables sums of 4 neighboring amplitudes were calculated establishing 5 new input parameters ($\Sigma_1$ -- $\Sigma_5$, $\Sigma_5$ is a sum of 3 amplitudes) -- see details in {}\cite{Panas:2018phd}. 
In order to avoid energy dependence, after the baseline subtraction, the pulses were normalized with respect to their energies.

For training and definition of the cuts, calibration data were used.
The Compton edge events of the 2615~keV peak (region between 2350~keV and 2375~keV) were chosen as SSEs (signal), and multiple Compton scattered events (region between 2450~keV and 2550~keV) as MSEs (background). 
The cut was defined requesting 80\% survival fraction for the DEP events. 
This allows to compare the present analysis with the corresponding one from the {\sc Gerda} Phase I data. 
No dedicated cut for high energy events was defined.

The stability of the input data was monitored using the time distribution of $\Sigma_2$. 
If instabilities (abrupt change of the input variable distribution) were observed, the data in the affected channels was divided into different (stable) sub-periods and analyzed separately. 
As a consequence, the cut for a given survival probability, was time dependent. 
Such a change for ANG5 has been observed between some runs, otherwise the distribution of $\Sigma_2$ was stable within less than 4\%.

The survival probability of the $0\nu\beta\beta$ decay events estimated from the simple cut based on the PL classifier is $(65.5 \pm 13.3)\%$, while the background in the region of interest is reduced by 56\%. Although, the signal efficiency is relatively low with conservatively estimated systematic uncertainty, PL allows also to eliminate efficiently the high energy events, about 87\% of them are rejected. 
An advantage of PL is the training performed only on $\gamma$-rays from the calibration runs.

This method was not retained for the final $0\nu\beta\beta$ decay search analysis due to lower signal efficiency and higher background in the region of interest.


\end{document}